\documentclass[aps, prd, preprintnumbers, floatfix, showpacs,10pt]{revtex4}

\usepackage[dvips]{graphics}
\usepackage{amsmath,amsfonts,bm}
\usepackage{epsfig,bm}
\usepackage{feynmf}
\usepackage{graphicx,comment}
\usepackage{dcolumn, color}

\def\cA{{\cal A}}

\def\tr{\mathop{\rm tr}}  \def\Tr{\mathop{\rm Tr}}  \def\Im{\mathop{\rm Im}}
\def\Re{\mathop{\rm Re}}

\def\pB{{\bar p}}

\def\beq{\begin{equation}}  
\def\eeq{\end{equation}}  
\def\bea{\begin{eqnarray}}  
\def\eea{\end{eqnarray}}

\newcommand{\be}{\begin{equation}}
\newcommand{\ee}{\end{equation}}
\newcommand{\bear}{\begin{eqnarray}}
\newcommand{\eear}{\end{eqnarray}}
\newcommand{\ba}{\begin{array}}
\newcommand{\ea}{\end{array}}

\newcommand{\dt}{\partial}

\begin{document}
\preprint{EFI 09-16}
\preprint{\today}


\title{The Pomeron contribution to $p ~p$ and $p ~ \bar p$ scattering in AdS/QCD}


\author{Sophia~K.~Domokos}
\affiliation{Enrico Fermi Institute and Department of Physics,
University of Chicago, Chicago Illinois 60637, USA}
\author{Jeffrey~A.~Harvey}
\affiliation{Enrico Fermi Institute and Department of Physics,
University of Chicago, Chicago Illinois 60637, USA}
\author{Nelia Mann}
\affiliation{Enrico Fermi Institute and Department of Physics,
University of Chicago, Chicago Illinois 60637, USA}


\begin{abstract}
We consider the differential and total cross sections for proton-proton and proton-antiproton scattering
in the Regge regime from the point of view of string dual models of QCD. We argue that the form factor which appears in the
differential cross section is related to the matrix element of the  stress tensor between proton states and give a procedure for computing the strength of the coupling of the Pomeron trajectory to the proton. We compute this coupling
in the Sakai-Sugimoto model and find excellent agreement with the data at large $s$ and small $t$.  The form factor can be estimated in the Skyrme model or in AdS/QCD models and gives a stiffer form factor than the commonly used electromagnetic form factor, in agreement with our fits to data.  Our model is also in good agreement with
the measured ratio of real to imaginary parts of the forward scattering amplitude at large $s$.

\end{abstract}



\maketitle


\section{Introduction}

In recent years, holographic QCD (hQCD) models based on the AdS/CFT correspondence have developed as a useful framework for understanding the structure of 
strongly coupled QCD. They have proved remarkably successful in reproducing meson masses, 
decay constants, and couplings. The difficulty of performing string calculations on curved backgrounds has largely 
limited their application to low-energy processes which can be studied in the supergravity limit of the string dual. However, such interesting
(and experimentally relevant) processes as the high-energy scattering of hadrons  lie outside this limit.
In particular, at large center of mass energy $\sqrt{s}$ and small momentum transfer $t$, perturbative QCD fails -- but considering the
exchange of only the lowest energy confined states in the supergravity limit of hQCD is also insufficient. Instead, one should
include the full tower of higher spin string excitations having the appropriate charge and parity. In this paper, we propose a technique inspired
by Regge theory and the structure of string scattering amplitudes to model the high $s$, low $t$ limit of proton-proton and
proton-anti-proton scattering.

The outline of this paper is as follows. In section II we review the connection between Regge theory and string theory
and discuss the AdS/QCD interpretation of the Pomeron. In section III we use  the structure of QCD string duals to develop a model 
for $p-p$ or $p- \bar p$ elastic scattering in the Regge limit. This model depends on four parameters: the slope and intercept
of the Pomeron trajectory, the strength of the coupling of the Pomeron to the proton, and a mass scale which determines
the relevant proton form factor. In section IV we compute the mass scale, the Pomeron-proton coupling, and 
the mass of the lowest particle on the trajectory, in the Sakai-Sugimoto model.
In section V we fit our model to data and compare the fit values of the parameters with the computed values.
In section VI we conclude and discuss some directions for future research.

\section{Regge theory, String theory, and the Pomeron}

In this section we  discuss some of the connections between Regge theory and string theory. While all of the results are known, we hope that this quick summary proves useful to string theorists unfamiliar with Regge theory  and to experts on Regge theory who may not have kept up with the latest developments in string theory and AdS/QCD. It will also serve to establish some of our key assumptions and to compare them with 
some of the phenomenological literature on Pomeron exchange. While Regge theory can certainly be  constructed independently of string theory, many of its results are more easily understood from a string-theoretic point of view; the Regge behavior of hadronic processes also hints strongly at the existence of a string dual description of  QCD.  As we will see, string duals of QCD illuminate some aspects of Regge theory which in the past were determined purely phenomenologically.

We are concerned in particular with the forward behavior of  proton-proton or
proton-antiproton scattering.  In the Regge limit of small $t$ and large $s$, the total cross section, the differential cross section, and the $\rho$ parameter are given in terms of the
invariant scattering amplitude ($\cA^{pp}(s,t)$ or $\cA^{p \bar p}(s,t)$) \footnote{For unpolarized scattering these quantities should be averaged over initial spins.} as
\beq
\sigma_{tot} = \frac{1}{s} \Im \cA(s,0), \qquad
\frac{d \sigma}{dt} = \frac{1}{16 \pi s^2} |\cA(s,t)|^2, \qquad
\rho(s) = \frac{\Re \cA(s,0)}{\Im \cA(s,0)} .
\eeq
QCD is strongly coupled in this regime, so a perturbative expansion in the QCD coupling cannot be used  to compute these quantities. It is natural, then, 
to try to compute them in the $1/N_c$ expansion, which often provides insight into the properties of QCD when standard perturbation theory fails \cite{Coleman,Manohar}.  At large $N_c$, the QCD spectrum includes infinite towers of narrow mesons and glueballs with arbitrarily high spin. For example, it is easy to construct gauge invariant operators with spin $J$ which have the same quantum numbers as the $\omega$
meson. These operators create higher mass and spin versions of the $\omega$, which also have nonzero
couplings to baryons.  Since the effective coupling between mesons is $g_{\rm eff} \sim 1/\sqrt{N_c}$, these states
become arbitrarily narrow as $N_c \rightarrow \infty$. Tree-level $t$-channel exchange of such a spin $J$ meson implies an amplitude which
behaves as $s^J$. Naively, terms with larger $J$ become more important at large $s$; the sum over
all such exchanges would then lead to amplitudes which grow rapidly at large $s$, violating the requirements of unitarity.

Regge theory instead sums up these exchanges of particles with fixed quantum numbers and higher mass and  spin (Regge trajectories) 
into a form that can be associated with a $t$-dependent pole at  $J= \alpha(t)$ in the complex angular momentum plane  \cite{Collinsbook}. There is a great deal of evidence that these Regge trajectories are well-approximated by a 
linear function, $\alpha(t) = \alpha(0) + \alpha' t$. At positive $t$ (corresponding to the crossed channel for a $2 \rightarrow 2 $ scattering process) this is evidenced by the fact that mesons can indeed be arranged into groups with $J = \alpha(0) + \alpha' M^2$.  For example, the trajectory which includes the
$I=1$ mesons $\rho, a_2, \rho_3, a_4$ has $\alpha(0) \sim 0.53$ and $\alpha' \sim 0.88 ~{\rm GeV}^{-2}$.  At small negative $t$, one can
extract $\alpha(t)$ from the differential cross section, which
in Regge theory has the characteristic form
\beq\label{regge_diffxsec}
\frac{d \sigma}{dt} = \beta(t) \left(\frac{s}{s_0}\right)^{2 \alpha(t) -2}~.
\eeq
Data from a variety of scattering processes shows that the linearity of trajectories extends to
moderate values of negative $t$ with the same slope and intercept as at positive $t$. 
While there is evidence that more complicated singularities are also required to describe the full range of hadronic data (e.g. Regge cuts) the basic picture of exchange of single Regge poles accounts for the structure of many hadronic scattering processes in the Regge limit.  

High-energy total cross section data quickly makes it clear, however, that a theory based purely on the exchange of known meson
states is not sufficient to describe the scattering of hadrons. The Regge behavior described above leads to total 
cross sections $\sigma_{tot}\propto (\Im \cA(s,0))/s$ which behave as
$s^{\alpha(0)-1}$ at large $s$. Since all the known meson Regge trajectories have intercepts $\alpha(0) \le 0.6$,
they cannot explain the experimental fact that total cross sections for hadronic processes tend to grow very slowly with
increasing $s$, at a rate independent of the quantum numbers of the scattered particles. This was apparent even in the early 1960's, and led 
\cite{Chew,Gribov}  to propose the existence of a new Regge
trajectory dubbed the ``Pomeron,'' with even signature, vacuum quantum numbers, and 
intercept $\alpha_{\cal P}(0) \sim 1$, that 
governed the large $s$ behavior of total cross sections. There have been many attempts over the years to fit total cross
sections with a combination of Reggeon and Pomeron contributions. See for example \cite{CGM,dl}, both of which conclude that the Pomeron has an intercept $\alpha_{\cal P} (0) \sim 1.06-1.08$ and a slope $\alpha'_{\cal P} \sim 0.25 ~{\rm GeV}^{-2}$.

However, the existence of the Pomeron trajectory and its structure has remained elusive for a number of reasons.
First of all, it is not easy to identify the corresponding trajectory at positive $t$, though many candidate states with the correct quantum numbers 
do exist.  It is not even clear that the Pomeron trajectory is unique. One could certainly imagine that there are 
several Pomeron trajectories, just as there are a variety of Reggeon trajectories. Second, the $s^{0.08}$ growth of total cross sections
implied by fits to the Pomeron intercept eventually  violates the Froissart-Martin bound, which requires that total
cross sections grow no faster than $\log^2 s$ as $s \rightarrow \infty$.  Thus single Pomeron exchange cannot be
the full story for sufficiently high $s$. Whether we have already reached a value of $s$ where such effects must be included is quite controversial. 
Attempts to test the question of single Pomeron exchange versus unitarized models of multi-Pomeron exchange
(or extrapolations of perturbative QCD using fits to forward data) have unfortunately 
led to inconclusive results \cite{Cudell}.  Finally, at increasing $|t|$ one encounters features
in the data which are not easily described by single Pomeron exchange with the same trajectory as at small $|t|$. This
has led to the idea of a ``hard Pomeron'' which in some papers is treated as a separate trajectory and in others is viewed as a change in the behavior of a single trajectory as $|t|$ increases. 

String dual models of QCD \cite{Wittenqcd,Sakai:2004cn,Erlich:2005qh,DaRold:2005zs} which grew out of the AdS/CFT correspondence \cite{Maldacena:1997re,Gubser,Witten} have shed some
light on these issues \cite{bpst}. In hQCD, meson states are dual to open strings and glueball states to closed strings. Analysis of the glueball spectrum in string duals \cite{glueballs} as well
as lattice gauge theory studies \cite{Meyer:2004jc} support the idea that there is a single Pomeron trajectory consisting
of glueball states, with the lowest mass state on the leading trajectory being a $2^{++}$ state and with the lowest mass $0^{++}$ state lying on a 
``daughter trajectory'' (a trajectory having the same quantum numbers as the ``leading'' trajectory, but a smaller intercept).  
The analysis of \cite{bpst} also establishes a connection between the
soft Pomeron and the hard BFKL Pomeron, which emerges from perturbative QCD \cite{bfkl}. For theories like QCD, which exhibit confinement and logarithmic running couplings, it leads to a theory where a discrete spectrum of poles at positive $t$  evolves continuously into a set of closely spaced poles on a flatter trajectory (see e.g. Fig. 11 in \cite{bpst}). The transition between these two behaviors is expected to occur at scales of order $\Lambda_{QCD}$. This picture very closely resembles one which emerges from models which generically use a 
fifth dimension, $r$, to encode the energy scale of the theory, and thus the running of the string tension \cite{Polyakov:1997tj}. The full
amplitude therefore becomes a sum over densely spaced trajectories with $\alpha(0)$ and $\alpha'$ depending on the energy scale $r$, so that a
different trajectory dominates for at each value of $t$. Analyzing the Veneziano and Virasoro-Shapiro amplitudes 
in this context (see e.g. \cite{Andreev:2004jm, Andreev:2004sy}) reveals a similar dependence of
the trajectory slopes to that found in \cite{bpst}.
Unfortunately,
the region of most interest for our analysis,  $- \Lambda_{QCD}^2 < t \le 0$, appears to be extremely model-dependent and
difficult to analyze in all such models.  
Based on the picture outlined above, we will assume that there is a single Pomeron trajectory and focus entirely on the regime in
which this trajectory promises to dominate.
We also assume that one 
of the primary effects of the curved space background in AdS duals is to shift the slope and the intercept of the leading closed string Regge trajectory from their flat space values. Computing this shift in the region of most phenomenological interest
is a difficult problem. For various approaches see \cite{bpst,Janik}.
We will later present an analysis of the effective trajectory extracted from data that suggests
a linear trajectory in the region $-0.6 ~{\rm GeV}^2 < t \le 0$, and will fit its slope and intercept.

We now review how the most important elements of Regge theory indeed emerge from string amplitudes. Here we consider
only flat space amplitudes, which are sufficient to illustrate Regge behavior.  In the next section we will assume
that amplitudes in curved space retain much of this structure but with shifted values for the parameters of the Regge trajectory.

Let us begin with an
amplitude for open string exchange, which should correspond to the exchange of Reggeons. The Veneziano amplitude, first introduced
as a model for $\pi+ \pi \rightarrow \omega + \pi$ scattering, and later recognized as defining open string scattering, can be written in terms of the amplitude
\beq
\cA^{\rm Ven}_{\{n,m,p\}}(s,t) =  \frac{\Gamma[n-a_o(s)] \Gamma[m-a_o(t)]}{\Gamma[p-a_o(s)-a_o(t)]}.
\eeq
The crossing symmetric combination appearing in Veneziano's original paper \cite{Veneziano} is
\beq
\cA^{\rm Ven}_{\{1,1,2\}}(s,t)+\cA^{\rm Ven}_{\{1,1,2\}}(t,u)+\cA^{\rm Ven}_{\{1,1,2\}}(s,u)~, 
\eeq
while the open bosonic string four-tachyon amplitude is simply
\beq
\cA_o(p_1,p_2,p_3,p_4) \propto  \cA^{\rm Ven}_{\{0,0,0\}}(s,t)+\cA^{\rm Ven}_{\{0,0,0\}}(u,t) + \cA^{\rm Ven}_{\{0,0,0\}}(s,u) 
\eeq
where we have ignored Chan-Paton factors and an overall constant.
In the above we take $a_o(x)$ to be
a linear function: $a_o(x)= a_o(0) + a'_o x$. To avoid notational confusion later on,  we use
$a_o(x)$ and $ a_c(x)$ for the linear functions which appear in open and closed string amplitudes and reserve
the notation $\alpha(0)$ and $\alpha'$ for the intercept and slope of the straight line relating $J$ to $M^2$ on a Regge
trajectory. Thus
for closed string Regge trajectories we have $J= \alpha_c(0) + \alpha'_c M^2$ and we define 
$\alpha_c(x)= \alpha_c(0) + \alpha'_c x$. This will become clearer in the following when we contrast
the Regge limits of open and closed string amplitudes. 


In the Regge limit ($s \rightarrow + \infty$ and  $t$ fixed), the Veneziano amplitude does not
behave smoothly: it has poles for large, positive, real $s$.  This can be remedied by giving $s$ a small imaginary part, which corresponds physically to
giving the meson states a small width, thus moving their poles slightly off the real axis. 
One finds that for large, complex $s$
\beq
\cA^{\rm Ven}_{\{0,0,0\}}(s,t) \rightarrow (-a'_o s)^{a_o(t)} \Gamma[-a_o(t)] = e^{-i \pi a_o(t)} (a'_o s)^{a_o(t)} 
\Gamma[-a_o(t)]  ,
\eeq
\beq
\cA^{\rm Ven}_{\{0,0,0\}}(u,t) \rightarrow (a'_o s)^{a_o(t)} \Gamma[-a_o(t)]  ,
\eeq
(using $u \sim -s$ in the Regge limit) and that $\cA^{Ven}_0(u,s)$, which does not have any t-channel poles,
vanishes exponentially in $s$ in this limit. From this we can easily see that the open string amplitude explicitly reproduces 
the Regge form of the differential cross section in (\ref{regge_diffxsec}).

The linear combinations of amplitudes which are even or odd under the exchange $u \rightarrow s$
have Regge limits
\bea
\cA_o^{+} \equiv \cA^{\rm Ven}_{\{0,0,0\}}(s,t)+\cA^{\rm Ven}_{\{0,0,0\}}(u,t)  \rightarrow  &(1 + e^{-i \pi a_o(t)}) (a'_o s)^{a_o(t)} \Gamma[-a_o(t)] \cr
\cA_o^{-} \equiv \cA^{\rm Ven}_{\{0,0,0\}}(s,t)- \cA^{\rm Ven}_{\{0,0,0\}}(u,t)  \rightarrow  &( -1 + e^{-i \pi a_o(t)})(a'_o s)^{a_o(t)} \Gamma[-a_o(t)] 
\eea
The prefactors in round brackets are known in Regge theory as ``signature factors'' and have zeroes when $a_o(t)$ is an odd
integer in $\cA_o^{+}$ or an even integer in $\cA_o^{-}$.
Using the product formula
\beq
\Gamma[-a_o(t)]= \frac{e^{\gamma a_o(t)}}{a_o(t)} \prod_{n=1}^\infty \frac{n e^{-a_o(t)/n}}{(a_o(t)-n)}
\eeq
we see that $\cA_o^{+}$ has poles at $a_o(t)=2k$ with residue $ \sim s^{2k}$  for $k$ a nonnegative integer.
Thus  $\cA_o^{+}$ corresponds to the exchange of a Regge trajectory of  particles with even  integer spin. Similarly,
$\cA_o^{-}$ corresponds to exchange of a Regge trajectory  with odd integer spin. Since in
either case we find the angular momentum of the particles being exchanged  near a t-channel
pole is $J= \alpha(t)$, for the open string we can identify $a'_o = \alpha'_o$ and $a_o(0)= \alpha_o(0)$ and
thus $a_o(t)= \alpha_o(t)$. When one considers the scattering of string states with higher spin, or extends the bosonic string to the superstring, the general structure of tree-level amplitudes in terms of ratios of Gamma functions remains the same, the only difference being 
the addition of a kinematic factor $K_o(1,2,3,4)$ which depends on the polarization tensors or spinor structures of the scattered particles and their
momenta. This implies that the higher mass and spin states being exchanged have couplings and vertex
factors tightly constrained by duality  in terms of the couplings and vertex factors of the lightest exchanged states.

We have now seen that the open (bosonic) string amplitude exhibits the correct pole and residue structure to reproduce the form of the
phenomenological Regge amplitude, and also naturally includes signature factors for exchange of resonances with even or odd spin.
Of course, the open bosonic string amplitude has obvious problems as a candidate for describing scattering of mesons. In particular, the ground state, corresponding to the first pole of $\cA_o^{+}$, is a tachyon and the first
spin one state exchanged in $\cA_o^{-}$ is a massless gauge field. However, one can easily modify
the string theory amplitude for phenomenological purposes  \cite{lovelace,shapiro} by using $\cA^{\rm Ven}_{\{1,1,1\}}$ instead of $\cA^{\rm Ven}_{\{0,0,0\}}$ and by using the phenomenologically determined values of the slope and intercept for meson trajectories.  

Let us now turn to closed string scattering, dual to the exchange of a trajectory of glueball states (i.e. the Pomeron). 
The simplest example is
the  scattering in flat space of four closed string tachyons, given by the Virasoro-Shapiro amplitude
\beq
\cA_c(p_1,p_2,p_3,p_4) = \frac{\Gamma[-a_c(s)] \Gamma[-a_c(t)] \Gamma[-a_c(u)]}{\Gamma[-a_c(t)-a_c(s)] \Gamma[-a_c(t)-a_c(u)] \Gamma[-a_c(u)-a_c(s)]} .
\eeq
When we use this amplitude for phenomenological purposes
we will include a kinematic prefactor $K_c(1,2,3,4)$ and fit the slope and intercept to data rather using the values from the critical bosonic string.

To take the Regge limit it is useful to write the $u$ dependence in terms of $s$ and $t$. For the simplest case
of $2 \rightarrow 2$ scattering of particles with equal mass $m$ we have
\beq \label{chieq}
a_c(s)+a_c(t)+a_c(u) = 4 a'_c m^2 + 3 a_c(0) \equiv \chi 
\eeq
so that 
\beq
\cA_c(p_1,p_2,p_3,p_4) = \frac{\Gamma[-a_c(s)] \Gamma[-a_c(t)] \Gamma[a_c(s)+a_c(t)- \chi]}{\Gamma[-a_c(t)-a_c(s)] \Gamma[a_c(t)- \chi] \Gamma[a_c(s)- \chi]} .
\eeq

Then, using the limits
\beq
\frac{\Gamma[-a_c(s)]}{\Gamma[-a_c(s)-a_c(t)]} \rightarrow (- a'_c s)^{a_c(t)} ,
\eeq
and
\beq
\frac{\Gamma[a_c(s)+ a_c(t) - \chi]}{\Gamma[a_c(s) - \chi]} \rightarrow (a'_c s)^{a_c(t)} ,
\eeq
we have the Regge limit of the Virasoro-Shapiro amplitude
\beq
\cA_c^{Reg}(p_1,p_2,p_3,p_4) = e^{-i \pi a_c(t)} (a'_c s)^{2 a_c(t)} \frac{\Gamma[-a_c(t)]}{\Gamma[a_c(t) - \chi]} .
\eeq
This amplitude has poles at $a_c(t)=n$ for $n=0,1,2, \cdots$ with residue $\sim s^{2n}$. This corresponds to t-channel exchange of a Regge trajectory with $J=2n=0,2,4, \cdots$. Note that it is not necessary  to do any projection or add amplitudes to get exchange of only even $J$ poles; this arises as a consequence of the fact that the leading
closed string Regge trajectory contains only even spin states. We also note that $J= 2 a_c(t)$, implying
that $\alpha'_c= 2 a'_c$ and $\alpha_c(0)= 2 a_c(0)$. 

The above discussion suggests two possible amplitudes to describe the exchange of an even signature 
Regge trajectory whose lowest state has spin $J_0=2$: (1) the Regge limit of either the even signature part of the Veneziano amplitude 
for open string theory or (2)  the Regge limit of the Virasoro-Shapiro amplitude for closed string theory. Rewritten
in terms of the actual Regge trajectories $\alpha_o(t)$ and $\alpha_c(t)$, including the kinematic prefactor, they are
\beq
\cA_o^{\rm Reg}(s,t) = [K_o(1,2,3,4)(\alpha'_o s)^{-2}] e^{-i \pi \alpha_o(t)/2} 2 \cos(\pi \alpha_o(t)/2) (\alpha'_o s)^{\alpha_o(t)} \Gamma(2-\alpha_o(t)),
\eeq
and
\beq
\cA_c^{\rm Reg}(s,t) = [K_c(1,2,3,4)(\alpha'_c s/2)^{-2}] e^{-i \pi \alpha_c(t)/2} (\alpha'_c s/2)^{\alpha_c(t)} \frac{\Gamma(1-\alpha_c(t)/2)}{\Gamma[\alpha_c(t)/2-1-\chi]} .
\eeq
By assumption the factor in square brackets approaches a constant at  large $s$.

The total cross section and the parameter $\rho = \Re \cA(s,0)/ \Im \cA(s,0)$ are identical in either case.
Provided the same Regge trajectory is used, they differ only by an
overall constant at $t=0$. However, the amplitudes do predict distinct differential cross
sections as these involve behavior at non-zero $t$. Regge amplitudes are often written in the form $(s/s_0)^{\alpha(t)}$
and we see that the reference scale $s_0$ differs in the two amplitudes. They also differ in their $t$ dependence
with the most dramatic difference being the existence of zeroes in the open string 
amplitude at $\alpha_o(t)= -1,-3,-5, \cdots$ (these are often called EXD zeroes in the 
literature and arise from the need to project out the even spin states from an
EXchange Degenerate trajectory with both even and odd spins) while the closed string amplitude has zeroes
coming from the poles in the denominator at $\alpha_c(t)= 2 \chi+2, 2 \chi, 2 \chi-2, \cdots $.

In the application of this formalism to proton-proton scattering discussed in the following section we use the
closed string amplitude to model Pomeron exchange, since the Pomeron trajectory is a closed string trajectory
in holographic duals of QCD. The situation is a bit murky, however. In dual models the proton is neither
an open string nor a closed string but rather a solitonic excitation of open strings (as expected from large $N_c$ reasoning
\cite{Wittenbaryon}), so while the exchanged states are indeed closed strings, the states being scattered are not.

\section{A general model for $p-p$ elastic scattering in the Regge regime}

We now use the behavior of the string amplitudes sketched above to develop a method for computing proton-proton scattering
in a holographic string dual of QCD. One should note that at present all the proposed duals have serious limitations; 
we also lack the technical tools to calculate the full tree-level string amplitudes in a curved space background, 
which should actually govern the behavior of proton-proton scattering. As a result we will have to make certain
approximations. Our first assumption (consistent with factorization in Regge theory) is that we can split the calculation into
two parts. (1) We determine the vertex which governs the coupling of the Pomeron to the
proton, which we assume is dictated by
the vertex for the lowest state on the Pomeron trajectory, the $2^{++}$ glueball. Using this vertex, we
compute the amplitude for tree-level exchange of a spin $2$ glueball. (2) We then convert this amplitude
into the Regge limit of a full tree-level string amplitude by ``Reggeizing''  the propagator. This is a heuristic
procedure, described in detail below,  which gives an answer consistent with the general principles of Regge theory. It should be a 
good starting approximation if the main effect of curved space and background fields on the string theory is to shift the slope and intercept 
of the Regge trajectory from their flat space values.

\subsection{The glueball coupling}

Our first task is to determine the coupling of the glueball field to the proton.  The glueball field can be treated as a second-rank symmetric traceless
tensor $h_{\mu \nu}$. Old ideas of tensor-meson dominance, \cite{freundtmd,tmdII} applied now
to the $2^{++}$ glueball rather than to the $f_2$ meson, suggest that
$h_{\mu \nu}$ should couple predominantly to the QCD stress tensor $T^{\mu \nu}$:
\beq
S_{\rm int}= \lambda \int d^4x h_{\mu \nu} T^{\mu \nu}.
\eeq
Assuming this is true, the glueball-proton-proton vertex is determined by  the matrix element of the stress tensor
between proton states
\beq
\langle p',s'| T_{\mu \nu}(0) | p, s \rangle ~.
\eeq
Using symmetry and conservation of $T_{\mu \nu}$ this matrix element can be written in terms of
three form factors \cite{Pagels} as
\beq\label{eq:EMTgeneral}
\langle p',s'| T_{\mu \nu}(0) | p,s \rangle=
\bar u(p',s') \biggl[ A(t) \frac{ \gamma_\alpha P_\beta+ \gamma_\beta P_\alpha}{2} + B(t)
\frac{i(P_\alpha \sigma_{\beta \rho} + P_\beta \sigma_{\alpha \rho}) k^\rho}{4m_p} + C(t)
\frac{(k_\alpha k_\beta - \eta_{\alpha \beta}k^2)}{m_p}   \biggr] u(p,s).
\eeq
where $k = p' - p'$, $t=k^2$ and $P = (p+p')/2$.
The fact that the proton has spin $1/2$ and mass $m_p$ implies the constraints $A(0)=1$ and $B(0)=0$.
We will see in the following subsection that the contribution from $C(t)$ is suppressed in the Regge limit
and that the contribution from $B(t)$ is small compared to that from $A(t)$. We note that gravitational form factors
of nucleons have been studied in the holographic context in \cite{Brodsky:2008pf} and also play an important role
in the analysis of deeply virtual Compton scattering \cite{Ji:1996nm}; for our purposes, however, a simple
analysis of their behavior in the Regge limit will suffice. 

How are the two ingredients -- the glueball and the proton -- manifest in a string dual description? 
Any hQCD model necessarily 
involves a theory containing gravity in a 5-dimensional (5d) space
(often along with additional compact directions), where one of the coordinates is dual to the energy
scale of QCD. There is therefore inevitably a graviton, which gives 
rise to a mode transforming as the desired $2^{++}$ glueball field.
In order to have dynamical mesons and baryons (rather than just a baryon vertex as in \cite{Wittenb})
there must also be fields in the theory which are dual to operators constructed out of quark fields. 
In particular, there must be a gauge field dual to  the axial-vector current operator which creates pions in QCD.  
In the large $N_c$ limit of QCD baryons may be treated as Skyrmions, that is as solitons of the pion field. 
Dual models lend themselves to this interpretation of baryons as Skyrmions, 
though
we could also contemplate adding in fields dual to baryon operators in QCD, as suggested in the recent work of \cite{Abidin:2009hr}.

Fluctuations $h_{MN}$ of the 5d background metric by definition
couple to the 5d stress tensor for the matter fields via
\beq
\frac{1}{2}\int d^5x \sqrt{g} ~T^{\rm matter}_{MN} h^{MN}
\eeq
where $T^{\rm matter}_{MN}$ includes a contribution from the fields which are dual to the pion. In this case,
$T_{MN}$ is the energy momentum tensor on a solitonic solution representing the proton. To reduce
this to a 4d coupling of the glueball  we expand the spin 2 piece of the metric perturbation $h^{\mu\nu}$ in terms of
the glueball wavefunction, and the matter fields in terms of the 4d pion and vector meson fields.  There is no guarantee
that this will yield a 4d coupling of the glueball predominantly to the 4d stress tensor of the proton.
However, analogy to similar results for vector meson dominance in dual models \cite{Hong,Grigoryan} lead us to believe
that this is indeed the case, and 
we show by explicit calculation that this is true in the  Sakai-Sugimoto model.
Making this assumption, then, we can evaluate the coupling $\lambda$ as an overlap integral of
pion and glueball wavefunctions.  

Given a semi-classical solution representing a baryon it is then straightforward following the discussion in
\cite{Cebulla:2007ei} to compute the relevant form factors.
We should note that we are computing this vertex in the large $N_c$ limit.
On the dual string theory side, this is a classical limit with the string coupling $g_s \rightarrow 0$, so
the calculation can be done in terms of a semi-classical solution to the equations of motion. 

\subsection{Tree level glueball exchange}

Let us now use the generic form of the vertex discussed in the previous section to calculate the 
cross section due to glueball exchange.  We will then ``Reggeize'' the glueball propagator wherever it appears
in the amplitude, to include in our model
scattering of higher spin glueballs.  

\begin{figure}
\includegraphics{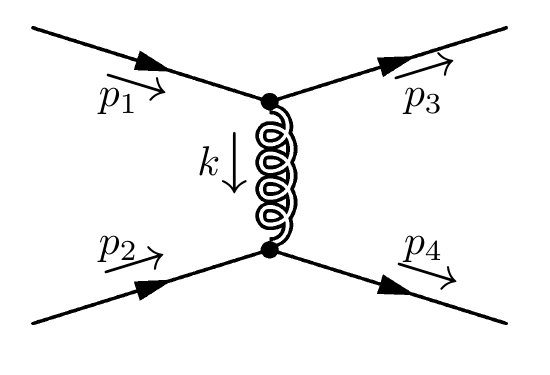}
\caption{\label{feyn} The Feynman diagram for the exchange of a glueball in the t-channel.}
\end{figure}

Consider a massive, spin-2 glueball exchanged in the $t$-channel (the dominant channel in the Regge limit).  
The Feynman diagram for this process is shown in FIG. \ref{feyn}.  

We use $p_1$ and $p_2$ as incoming momenta and $p_3$ and $p_4$ as outgoing momenta, with $s$, $t$, and $u$ defined in the usual way, with
 $k = p_1 - p_3 = p_2 - p_4$ the momentum of the glueball.
The massive spin-2 propagator (as given in \cite{prop}) is
\beq
\frac{d_{\alpha\beta\gamma\delta}(k)}{k^2 - m_g^2}
\eeq
where $m_g$ is the mass of the glueball and the indices contracted at one side of the propagator are $\alpha,\beta$, the indices at the other end are $\gamma,\delta$, and
\beq
d_{\alpha\beta\gamma\delta} = \frac{1}{2}(\eta_{\alpha\gamma}\eta_{\beta\delta} + \eta_{\alpha\delta}\eta_{\beta\gamma}) - \frac{1}{2m_g^2}(k_{\alpha}k_{\delta}\eta_{\beta\gamma} + k_{\alpha}k_{\gamma}\eta_{\beta\delta} + k_{\beta}k_{\delta}\eta_{\alpha\gamma} + k_{\beta}k_{\gamma}\eta_{\alpha\delta})
\eeq
$$
 + \frac{1}{24}\left[\left(\frac{k^2}{m_g^2}\right)^2 - 3\left(\frac{k^2}{m_g^2}\right) - 6\right]\eta_{\alpha\beta}\eta_{\gamma\delta} - \frac{k^2 - 3m_g^2}{6m_g^4}(k_{\alpha}k_{\beta}\eta_{\gamma\delta} + k_{\gamma}k_{\delta}\eta_{\alpha\beta}) + \frac{2k_{\alpha}k_{\beta}k_{\gamma}k_{\delta}}{3m_g^4}.
$$
Using the glueball-proton-proton coupling in the form of eq. (\ref{eq:EMTgeneral}) from the previous section, the amplitude becomes
\beq
\cA = \frac{\lambda^2d_{\alpha\beta\gamma\delta}}{4(t - m_g^2)}\Big[A(t)(\bar{u}_{1}\gamma^{\alpha}u_{3})(p_1 + p_3)^{\beta} + \frac{iB(t)}{2m_p}(p_1 + p_3)^{\beta}k_{\rho}(\bar{u}_{1}\sigma^{\alpha\rho}u_{3}) + \frac{C(t)}{m_p}(\bar{u}_1 u_{3})(k^{\alpha}k^{\beta} - \eta^{\alpha\beta}t)\Big]
\eeq
\beq
\nonumber
\times \Big[A(t)(\bar{u}_{2}\gamma^{\gamma}u_{4})(p_2 + p_4)^{\delta} + \frac{iB(t)}{2m_p}(p_2 + p_4)^{\delta}k_{\lambda}(\bar{u}_{2}\sigma^{\gamma\lambda}u_{4}) + \frac{C(t)}{m_p}(\bar{u}_2 u_{4})(k^{\gamma}k^{\delta} - \eta^{\gamma\delta}t)\Big]~.
\eeq
Using the Dirac equation
we can see that
\beq
k_{\mu}(\bar{u}_1\gamma^{\mu}u_{3}) = k_{\mu}(\bar{u}_2\gamma^{\mu}u_{4}) = 0
\eeq
and
\beq
k_{\mu}(p_{1} + p_{3})^{\mu} = k_{\mu}(p_{2} + p_{3})^{\mu} = 0.
\eeq
Furthermore, the second structure in the vertex will vanish when dotted into pairs of $k^{\mu}$.  This means we can ignore all terms in the propagator except those of the form $\eta\eta$.  We will be taking the Regge limit of the amplitude, so we can drop any terms that will be suppressed by factors of $t/s$ or $m_p^2/s$.  The amplitude then becomes
\beq
\cA = \frac{\lambda^2}{8(t - m_g^2)}\left\{2sA^2(t)(\bar{u}_1 \gamma^{\alpha} u_3)(\bar{u}_2 \gamma_{\alpha} u_4) + 4A^2(t)p_2^{\alpha}p_1^{\beta}(\bar{u}_1 \gamma_{\alpha} u_3)(\bar{u}_2 \gamma_{\beta} u_4) + \frac{iA(t)B(t)}{2m_p}\left[2s k^{\rho} (\bar{u}_1\sigma_{\alpha\rho} u_3)(\bar{u}_2 \gamma^{\alpha} u_4) \right. \right.
\eeq
\beq
\nonumber
\left.
 + 2s k_{\lambda}(\bar{u}_1 \gamma_{\gamma} u_3)(\bar{u}_2\sigma^{\gamma\lambda} u_4) + 4p_{2\rho}p_{4\alpha}p_1^{\beta}(\bar{u}_1 \sigma^{\alpha\rho} u_3)(\bar{u}_2\gamma_{\beta} u_4) + 4p_{1\lambda }p_{3\gamma}p_2^{\beta}(\bar{u}_1 \gamma_{\beta} u_3)(\bar{u}_2 \sigma^{\gamma\lambda} u_4) \right]
\eeq
\beq
\nonumber
- \frac{B^2(t)}{4m_p^2}\left[2s k_{\rho}k^{\lambda}(\bar{u}_1\sigma^{\alpha\rho}u_3)(\bar{u}_2\sigma_{\alpha\lambda}u_4) + 4p_{2\rho}p_{4\alpha}p_{1\lambda}p_{3\gamma}(\bar{u}_1 \sigma^{\alpha\rho} u_3)(\bar{u}_2 \sigma^{\gamma\lambda} u_4)\right]
\eeq
\beq 
\nonumber
\left. + \left[\frac{1}{12}\left[\left(\frac{t}{m_g^2}\right)^2 - \frac{3t}{m_g^2} - 6\right]\left(2m_p A(t) + \frac{B(t)t}{2m_p} - \frac{3C(t)t}{m_p}\right)^2 + \frac{2tC(t)}{m_p^2}\left(3C(t) - tB(t) - 4m_p^2A(t)\right)\right](\bar{u}_1u_3)(\bar{u}_2u_4)\right\}.
\eeq
The cross section can then be calculated from the amplitude, and it will be proportional to
\beq
\frac{1}{4}\sum_{\mathrm{spins}} |\cA|^2= \frac{\lambda^4 s^4}{(t - m_g^2)^2}\left(A^2(t) - \frac{t B^2(t)}{16 m_p^2}\right)^2 + \cdots
\eeq
where we suppress all terms subleading in $t/s$ or $m_p^2/s$.  

The form factor $B(t)$ is zero at $t = 0$ and slowly varying. Together with the factor of $\frac{t}{16 m_p^2}$ this implies that at small $t$,
the term proportional to $B^2(t)$ is very small compared to the $A^2(t)$ term.  The part of the cross section proportional to $A(t)$ dominates, allowing us to drop all other terms and simply associate a $\lambda A(t)$ to each vertex in the amplitude, giving
\beq\label{eq:glueballdxc}
\frac{d\sigma}{dt} = \frac{\lambda^4 s^2A^4(t)}{16\pi(t - m_g^2)^2}
\eeq
as the cross section for spin 2 glueball exchange. Note that the appearance of $s^2$ in the numerator is precisely the correct $s^J$ dependence
expected from the exchange of a spin-2 particle. The denominator $(t - m_g^2)^2$ will be replaced by the square of the ``Reggeized'' propagator we describe
in the next subsection.

\subsection{Reggeizing the propagator}

Having computed the cross section for exchanging the lightest state on the $2^{++}$ trajectory, we must include
the higher spin states on the trajectory, which correspond to stringy excitations on the curved background. As noted above, the computation of the full string amplitude is prohibitively difficult. Instead, we analyze in greater detail the scattering amplitude for four closed strings in flat space, and from this extract the propagator for a $t$-channel closed string exchanged in the Regge limit.

As discussed in the previous section, the scattering amplitudes for closed bosonic strings and for closed superstrings take the form
\beq
\cA = \frac{\Gamma[-a(t)]\Gamma[-a_c(u)]\Gamma[-a_c(s)]}{\Gamma[-a_c(t) - a_c(s)]\Gamma[-a_c(t) - a_c(u)]\Gamma[-a_c(u) - a_c(s)]}K_c(p_1, p_2, \dots)
\eeq
where $K_c$ is a kinematic factor with no poles, which depends on the momenta and polarizations of the scattered 
particles. $a_c(x)$ is some linear function related to the spectrum of closed strings:
\beq
a_c(x) = a_c(0) + a_c'x.
\eeq
We assume that this basic form of the amplitude holds true in (weakly) curved space, with $a_c(0)$, $a_c'$ and the kinematic factor $K_c$ undetermined and 
dependent on the details of the geometrical background.  This amplitude boasts many features that are important to modeling Pomeron exchange. For example, it is completely symmetric under the exchanges of $s, t$, and $u$.  We know experimentally that proton/proton scattering and proton/anti-proton scattering have the same behavior in the limit where Pomeron exchange dominates, and we know that crossing symmetry therefore requires that the amplitude be completely symmetric under exchanges of $s, t$, and $u$.  In addition, the amplitude has the correct pole and residue structure to describe the exchange of even spin, vacuum quantum number
states.

To find the proper ``Reggeized'' replacement for the spin 2 glueball propagator in (\ref{eq:glueballdxc}), we first expand 
the amplitude around one of the $t$-channel poles, which occur at $a_c(t) = n$, with $n = 0, 1, 2, \dots$:
\beq
\cA \approx \frac{(1+a_c(s))\cdots(n+a_c(s))(-\chi + a_c(s))\cdots(-\chi + n - 1 + a_c(s))}{n!\Gamma[n - \chi](n - a_c(t))}K_c(p_1, p_2, \dots)
\eeq
where $\chi$ is defined as in eq. (\ref{chieq}) and $K_c$ is evaluated at $a_c(t) = n$.  As we would expect for the sum of  exchanges of all particles on the Regge trajectory,  the residue of the pole is a polynomial in $s$,
and we can identify the leading behavior of this polynomial as $s^{J}$ if the particle being exchanged has  spin $J$.  That is
\beq
\cA \approx \frac{P_{2n}(a_c's) (s^{k}f(t, \epsilon_i) + \cdots)}{n!\Gamma[n - \chi](n - a_c(t))}
\eeq
where 
\beq
K_c(p_1, p_2, \dots) \propto s^{k}f(\epsilon_i) + \cdots
\eeq
with $f(\epsilon_i)$ some unknown function of the polarizations of the scattered particles which will factor out of our calculation in the end. $P_{2n}(a_c's)$ is a polynomial of degree $2n$ whose first term is $(a_c's)^{2n}$.  The spin of the $n$th particle on the trajectory is then
\beq
J = k + 2n.
\eeq
We can therefore use the leading $s$ behavior of the kinematic factor $K_c$ to arrange the spin of the 
lowest particle on the trajectory to be whatever we want, after which the higher spins are completely determined.  
In our case, the $n = 0$ pole should correspond to a spin-2 particle, meaning that $k = 2$.  This
 implies that the trajectory of particles contributing to the amplitude in the Regge limit (where $s$ is large) 
 consists only of even spin particles, consistent with the Pomeron coupling identically to particles and their anti-particles.  When we assume that the Pomeron is the dual of a closed string, this comes about quite naturally.  By contrast, if we had assumed that the Pomeron is 
 dual to an open string, we would have used the Veneziano amplitude,
  which has poles for both even and odd spin particles.
 
Let us now now relate our string amplitude parameters $a_c(0)$ and $a_c'$ to the traditional parameters of Regge theory.  If we have
\beq
\alpha_c(0) + \alpha_c' m_{J}^2 = J
\eeq
and $J = 2n + 2$, then we need
\beq
2a_c(0) + 2 = \alpha_c(0), \hspace{.5in} \mathrm{and} \hspace{.5in} 2a_c' = \alpha_c'.
\eeq

In the Regge limit (where we keep only the leading $s$ behavior), the amplitude from the exchange of the lowest particle on the trajectory will simply be
\beq
\label{lowest}
\cA \approx \frac{- s^{2}f(\epsilon_i)}{a_c'\Gamma[-\chi](t - m_g^2)}
\eeq
where we have used the fact that the mass of the lowest particle on the trajectory (the spin-2 particle) is
\beq
m_g^2 = m_2^2 = -\frac{a_c(0)}{a'_c}.
\eeq

If we take the Regge limit of the full amplitude, however, we find
\beq\label{eq:dhm_amp}
\cA \rightarrow \frac{\Gamma\left[1 - \frac{\alpha_c(t)}{2}\right]}{\Gamma\left[\frac{\alpha_c(t)}{2} -1 - \chi\right]}e^{-i\pi\alpha_c(t)/2}\left(\frac{\alpha'_c s}{2}\right)^{\alpha_c(t) - 2} s^{2}f(\epsilon_i)
\eeq
(where we are now using the characteristic parameters $\alpha_c(0), \ \alpha_c'$ of the Regge trajectory).  Note again that the factor $f(\epsilon_i)$ contains
all information about the incoming and outgoing particles. We can thus relate this amplitude to the amplitude in eq. (\ref{lowest}) by
replacing the glueball exchange factor $\frac{1}{t - m_g^2}$ with a Reggeized Pomeron propagator
\beq
\frac{1}{t - m_g^2} \rightarrow \frac{-\alpha_c'}{2}\frac{\Gamma[-\chi]\Gamma\left[1 - \frac{\alpha_c(t)}{2}\right]}{\Gamma\left[\frac{\alpha_c(t)}{2} -1 - \chi\right]}e^{-i\pi\alpha_c(t)/2}\left(\frac{\alpha'_c s}{2}\right)^{\alpha_c(t) - 2}~.
\eeq
The factors of $f(\epsilon_i)$ have indeed cancelled. We can now find the full Pomeron contribution to proton/proton scattering by applying this same replacement rule to the graviton propagator in (\ref{eq:glueballdxc}).  The proton/proton differential cross section becomes
\beq\label{eq:dhm_diffxsec}
\frac{d\sigma}{dt} = \frac{\lambda^4 A^4(t) \Gamma^2[-\chi]\Gamma^2\left[1 - \frac{\alpha_c(t)}{2}\right]}{16\pi \Gamma^2\left[\frac{\alpha_c(t)}{2} - 1 - \chi\right]} \left(\frac{\alpha'_c s}{2}\right)^{2\alpha_c(t) - 2},
\eeq
corresponding to the invariant amplitude
\beq\label{eq:dhminvamp}
\cA(s,t)= s \lambda^2 A^2(t) e^{-i \pi \alpha_c(t)/2} \biggl( \frac{\Gamma[-\chi]\Gamma\left[1 - \frac{\alpha_c(t)}{2}\right]}{\Gamma\left[\frac{\alpha_c(t)}{2} -1 - \chi\right]} \biggr) \left(\frac{\alpha'_c s}{2}\right)^{\alpha_c(t) - 1}.
\eeq

This form provides a model for the differential cross section, total cross section and $\rho$ parameter of either proton/proton or proton/anti-proton 
scattering at very high center of mass energy, where the process is dominated by Pomeron exchange. 
This prediction is relatively model-independent, relying only on the structure
of the closed string amplitude and the assumption that the graviton (which must be present in any dual theory) couples
to the energy-momentum tensor.
 It depends on
 four parameters:  the two trajectory parameters $\alpha_c(0)$ and $\alpha'_c$,  the glueball-proton-proton coupling strength $\lambda$, 
 and the dipole mass $M_d$ (for small $t$ we can approximate the form factor $A(t)$ with a
dipole $A(t)=(1-t/M_d^2)^{-2}$). 
 Of these unknowns, the coupling $\lambda$, the dipole mass $M_{d}$, and the glueball mass $m^2_{g} = (2 - \alpha_c(0))/\alpha_c'$ are all present in the low-energy process involving the exchange of the lowest glueball.  That is, if we know the low-energy process, which  we can compute in the
 supergravity limit, then the \emph{only} dependence on the full string theory lies in determining the trajectory slope $\alpha'_c$.
 
We now present two approaches for fixing the four parameters of the model: (1) we calculate three of them in
a specific dual model, and (2) we compare these results to least-squares fits of our model to scattering data.

\section{Computation of parameters in the Sakai-Sugimoto model}

After briefly reviewing the Sakai-Sugimoto model \cite{Sakai:2004cn}, we compute the parameters
$m_g$, $\lambda$, and $M_d$. 
We note first that the Sakai-Sugimoto model depends on three quantities, $M_{KK}$, $g_{YM}$, and $l_s$.  The string scale $l_s$ does not appear in the low-energy supergravity limit, only in the full string theory. The other two parameters
are arbitrary a priori, but they
may be fitted using the $\rho$ mass and the pion decay constant \cite{Sakai:2004cn}.  We find
excellent agreement between $m_g$, $M_d$, and $\lambda$ as computed in the fully fixed Sakai-Sugimoto model, and the values determined by fitting to 
$pp$ and $p\overline{p}$ scattering data.

\subsection{Sakai-Sugimoto Model}

The Sakai-Sugimoto model \cite{Sakai:2004cn, Sakai:2005yt}, is a top-down QCD dual: it relies on a brane construction in
10d supergravity to produce the salient features of strongly coupled QCD, such as confinement and chiral symmetry breaking.
$N_f$ flavor $D8$-branes are placed in the background generated by $N_c$ $D4$-branes.  Closed string (or bulk) excitations are dual to QCD glueballs, 
while open strings living on the probe branes and transforming in the adjoint of a $U(N_f)$ symmetry are dual to scalar and vector mesons.

The $D4$ background is defined by the following metric, dilaton, and Ramond-Ramond three-form $C_3$ (with $F_4=dC_3$):
\begin{align}\label{eq:D4sugra}
ds^2=\left(\frac{U}{R}\right)^{3/2}\left(\eta_{\mu\nu}dx^{\mu}dx^{\nu} + f(U)d\tau^2\right)
+ \left(\frac{R}{U}\right)^{3/2}\left(\frac{dU^2}{f(U)} + U^2 d\Omega_4^2 \right) \ , \\[5pt] \nonumber
e^\phi = g_s\left(\frac{U}{R}\right)^{3/4} \ , \ \quad F_4 = \frac{2\pi N_c}{V_4}\epsilon_4 \ , \ \quad f(U) \equiv 1 - \frac{U_{KK}^3}{U^3} \ .
\end{align}
A radial coordinate $U$ and
a unit $S^4$ parametrize the directions transverse to the $D4$-branes. 
 $d\Omega_4^2$ is the metric on the unit $S^4$, which has volume form $\epsilon_4$ and volume $V_4=8\pi^2/3$. $R^3=\pi g_sN_cl_s^3$.
The $D4$-branes are extended in the $\tau$ and $x^\mu$-directions, for $\mu=0,1,2,3$.  The $\tau$-direction is 
made periodic with  $\tau\sim\tau+ 2\pi M_{KK}^{-1}$. The radial coordinate $U$ must now be
 bounded from below ($U\ge U_{KK}$) to avoid a conical singularity. 
 In order to break the remaining supersymmetry, we impose
 anti-periodic boundary conditions on the fermionic modes
so they acquire masses of order $M_{KK}$.

 It is often useful to work with the scale and the effective 4d coupling constant 
 of the Yang-Mills theory. 
 In terms of $U_{KK}$ and $R$, $M_{KK}=3U_{KK}^{1/2}/(2R^{3/2})$, and $ g^2_{YM} = 2\pi M_{KK} g_s l_s $.

Placing $N_f$ $D8$ branes in this background produces flavor degrees of freedom. In the probe limit ($N_f\ll N_c$) the 
backreaction of the $D8$-branes with the $D4$-brane geometry is negligible. The flavor branes assume a nontrivial profile in the $(U,\tau)$
plane, and are fully extended along the $x^\mu$ and the $S^4$ directions.

\subsection{Open and closed string spectra}
Mass spectra of excitations coming from bulk and brane modes can be determined by perturbing the supergravity 
and the brane (DBI) actions, respectively.  Computations of the glueball (i.e. closed string) spectrum for this background geometry were performed in  \cite{Brower:2000rp} and \cite{Constable:1999gb}. We briefly review the treatment of \cite{Brower:2000rp} using the conventions of \cite{Sakai:2004cn}, 
and cite relevant results.

The 10d bulk field content consists of the graviton, $h_{MN}$, a dilaton $\phi$, an NS-NS tensor
$B_{MN}$, and R-R one- and three-forms $C_M$ and $C_{MNL}$. Neglecting any dependence on the $S^4$ transverse to
the $D4$-branes, and ignoring all but the lowest $KK$ modes in the compactified $\tau$ direction essentially reduces the problem to
five dimensions: $(x^\mu, U)$. We can classify the states according to their transformation properties under $SO(3)$ rotations
in the physical space directions $(x_1,x_2,x_3)$ of the field theory. The graviton in particular gives rise 
to a $2^{++}$ state $(h_{ij})$, a $1^{-+}$
state $(h_{i\tau})$, and a $0^{++}$ state $(h_{\tau\tau})$. The coupling of the bulk fields to the boundary gauge theory and 
the parity- and charge-conjugation invariance of
the overall action determine the parity and charge quantum numbers of the 4d field theory states.

Now consider the standard supergravity action 
\begin{equation}
S=\frac{1}{2\kappa_{10}^2}\int d^{10}x\sqrt{-G}\left\{ e^{-2\phi}(R+4(\nabla\phi)^2 )-\frac{(2\pi)^4l_s^6}{2\cdot 4!}F_4^2 \right\}
\end{equation}
where $\kappa_{10}$ is the 10d Newton constant. We introduce perturbations around the background
metric by taking $G_{MN}=\widetilde{G}_{MN}-h_{MN}$, where $\widetilde{G}_{MN}$ is the background metric from eq. (\ref{eq:D4sugra}). 
Varying with respect to $h_{ij}$ ($i,j$ are spatial Lorentz indices), we have the equations of motion
\begin{equation}
-\frac{1}{2}\left( \frac{9f}{2}+3Uf'\right)h_{ij}+\left( f+Uf'\right) U \dt_U h_{ij}+f U^2\dt_U^2h_{ij}=-\frac{q^2 R^3}{U}h_{ij}~.
\end{equation}
The prime denotes differentiation with respect to $U$, and  $q^2$ is the 4d momentum of the mode with $h_{ij}(q,U)=\int d^4x e^{-iqx}h_{ij}(x,U)$.  We work in a gauge where $h_{0M}=0$, $\dt_ih_{AB}=0$ and retain the traceless (spin 2) 
piece of $h_{ij}$.  Writing the 10d perturbation $h_{ij}(x,U)$ as a tower of resonances,
 \begin{equation}\label{eq:hexp}
 h_{ij}(x,U)=\sum\limits_{n=1}^\infty h^{(n)}_{ij}(x)\left(\frac{U}{R}\right)^{3/2}T_n(U)~,
 \end{equation}
the equation of motion becomes an eigenvalue equation for the modes $T_n(U)$ with $q^2=m_n^2$, where $m_n$ is the
mass of the $n^{th}$ resonance:
\begin{equation}
\dt_U\left( U^4f\dt_UT_n\right) =-m_n^2R^3UT_n~.
\end{equation}
As discussed earlier, we work in a strict Regge limit, only taking into account contributions from the
leading Regge trajectory, and not from the daughter trajectories. We will therefore use only the lightest mode in the KK tower
of $2^{++}$ glueballs. For simplicity, we define $T(U)\equiv T_n(U)$, $h_{ij}\equiv h^{(1)}_{ij}$, and 
$m_g^2\equiv m_1^2$ 
\footnote{Note that there is an additional KK tower due to excitations in the compact $\tau$ direction. Again, we are only interested in the lightest mode, 
so we assume that $h_{ij}$ also has zero momentum in the $\tau$ direction. } .
The mass of the lightest glueball is proportional to the lowest eigenvalue,  
\begin{equation}\label{eq:gluemass}
m_g^2=1.57~M_{KK}^2~,
\end{equation} 
which agrees with the result derived
in \cite{Brower:2000rp}. 

Having computed the mass of the lightest spin 2 graviton mode, we must now normalize its wavefunction to yield
a canonical kinetic term in the effective 4d action. 
The graviton kinetic term comes from
\begin{equation}
S = \frac{1}{2\kappa_{10}^2}\int d^{10}x \sqrt{-G} e^{-2\phi} R
\end{equation}
which we expand to quadratic order in $h_{\mu\nu}$:
\begin{equation}
\sqrt{-G} e^{-2\phi} R = \frac{1}{2}\sqrt{-\tilde{G}}e^{-2\tilde{\phi}}\tilde{G}^{\mu\nu}\tilde{G}^{\beta\delta}\tilde{G}^{\alpha\gamma}\left[\partial_{\alpha}h_{\delta\nu}\partial_{\mu}h_{\beta\gamma} - \frac{1}{2}\partial_{\alpha}h_{\delta\nu}\partial_{\gamma}h_{\beta\mu}\right] + \cdots ~,
\end{equation}
assuming again that the graviton is traceless.  Using the expansion in (\ref{eq:hexp}) and writing the integral in terms of the dimensionless ratio $U/U_{KK}$,
\begin{equation}
S = \frac{N_c^3 M_{KK}^2 g_{YM}^2}{3^5\pi^2} \int_{1}^{\infty} \frac{U}{U_{KK}} T^2(U)d\left(\frac{U}{U_{KK}}\right) \int d^{4}x \frac{1}{2}\eta^{\mu\nu}\eta^{\beta\delta}\eta^{\alpha\gamma}\left[\partial_{\alpha}h_{\delta\nu}\partial_{\mu}h_{\beta\gamma} - \frac{1}{2}\partial_{\alpha}h_{\delta\nu}\partial_{\gamma}h_{\beta\mu}\right] + \cdots~.
\end{equation}
The coefficient of the kinetic term for the 4d spin 2 mode is 
\begin{equation}
{\cal N}_T^2=\frac{N_c^3 M_{KK}^2 g_{YM}^2}{3^5\pi^2} \int_{1}^{\infty} \frac{U}{U_{KK}} T^2(U)d\left(\frac{U}{U_{KK}}\right)\equiv \frac{N_c^3 M_{KK}^2 g_{YM}^2}{3^5\pi^2}{\cal I_T}~,
\end{equation}
where $T(U)$ has dimensions of length and ${\cal I}_T$ has dimensions of length squared.
Rescaling $T(U)\rightarrow T(U)/\cal{N}_T$ yields a canonically normalized kinetic term.

Now we turn to the open string spectrum on the $D8$-branes, given by the leading DBI action
\begin{equation}\label{eq:fullDBI}
S_{D8}^{DBI}=-T_8\int d^9x e^{-\phi}\tr\sqrt{-\det \left( G_{MN}+2\pi\alpha' F_{MN}\right)} ~
\end{equation} 
where $T_8=(2\pi)^{-8}l_s^{-9}$ is the brane tension, $G_{MN}$ is the pull-back of the background metric onto
the brane stack, and $F_{MN}=\dt_M A_N-\dt_N A_M-i[A_M,A_N]$ is the non-abelian field strength of the $U(N_f)$ gauge fields living
on the branes. We have used the  convention $\tr (T^aT^b)=\delta^{ab}/2$ for the gauge group generators. 

By extremizing the DBI action without gauge field fluctuations ($A_M=0$), we can determine the profile of
the $D8$-brane stack in the $(U,\tau)$ plane:
\begin{equation}
\tau(U)=\pm U_0^4 f(U_0)^{1/2}\int_{U_0}^U\frac{dU}{\left(\frac{U}{R}\right)^{3/2}f(U)\sqrt{U^8 f(U)-U_0^8f(U_0)}}~,
\end{equation}
where $U_0$ is a constant of integration. The geometry of the branes explicitly realizes chiral symmetry breaking. 
The radial coordinate $U$ corresponds to the energy
scale of the dual field theory. Near the UV boundary ($U\rightarrow\infty$), the solution exhibits chiral symmetry: it resembles a pair of  
parallel $D8$- and $\overline{D8}$ stacks separated
in $\tau$ by some distance $L$, with 4d modes transforming under $U(N_f)_L\times U(N_f)_R$. 
As $U$ (the energy scale) decreases, the $D8$ and $\overline{D8}$ branes curve toward each other, until
they meet at $U=U_0$, breaking the chiral symmetry of the two independent UV brane stacks to $U(N_f)_V$. 
Like \cite{Sakai:2004cn} , we focus on the solution where $U_0=U_{KK}$, and the $D8$ and $\overline{D8}$ lie at antipodal points
on the $\tau$ circle. It will prove convenient to parametrize the direction along the probe branes in $(U,\tau)$ plane with 
\begin{equation}
Z=\pm \left[\left(\frac{U}{U_{KK}}\right)^3-1\right]^{1/2}~, 
\end{equation}
where $Z\in (-\infty,\infty)$ such that $Z\rightarrow\pm\infty$ are the left (right) UV boundaries.

The gauge field fluctuations in the $(\mu,Z)$ directions on the $D8$-brane give rise to towers of 4d vector and axial-vector meson states in
the field theory.
Assuming no dependence on the $S^4$ coordinates, the DBI action (\ref{eq:fullDBI}) becomes
\begin{equation}
S_{D8} = -\kappa \int  d^4x dZ \tr\left[ \frac{1}{2}K(Z)^{-1/3}\eta^{\mu\nu}\eta^{\rho\sigma}F_{\mu\rho}F_{\nu\sigma}+M_{KK}^2 K(Z)\eta^{\mu\nu}F_{\mu Z}F_{\nu Z}+\dots\right]~,
\end{equation}
with
\begin{equation}
\kappa = \frac{g_{YM}^2N_c^2}{216\pi^3}\quad\text{and}\quad K(Z)=1+Z^2~.
\end{equation}

In order to ensure that the mass and kinetic terms of the $4d$ action are normalizable, we must have the field strengths $(F_{\mu Z},~F_{\mu\nu})\rightarrow 0$ as $Z\rightarrow\pm\infty$. 
We can choose a gauge where $A_M$ itself vanishes at large $Z$. 
In order to more conveniently realize the (axial)vector meson
spectrum and the chiral Lagrangian for the pion modes, we follow \cite{Sakai:2004cn} and transform to a gauge where $A_Z=0$ :
\begin{align}\label{eq:gaugetransformedA}
A_\mu(x^\mu,Z)&\rightarrow A_\mu^{(g)}=gA_\mu g^{-1}+ig\dt_\mu g^{-1} \\
A_Z(x^\mu,Z)&\rightarrow 0~,
\end{align}
where the gauge transformation $g$ has the form of a Wilson line,
\begin{equation}
g(x^\mu,Z)=P\exp\left[ -i\int_0^Z dZ' A_Z(x^\mu,Z)\right]~.
\end{equation}
$A_\mu^{(g)}$ now splits naturally into a normalizable piece, $gA_\mu g^{-1}$, which gives rise to the (axial)vector meson spectrum, and
a nonnormalizable piece $ig\dt_\mu g^{-1}$. 

Focusing on the non-normalizable modes in the DBI action, we arrive at the action of the 4d Skryme model whose solitonic excitations
we identify with baryons. 
The non-normalizable piece of $A_\mu^{(g)}$, $ig\dt_\mu g^{-1}$, changes the boundary conditions on the gauge field such that 
\begin{equation}
\lim\limits_{Z\rightarrow\pm\infty}A_\mu^{(g)}(x^\mu,\pm\infty)= \lim\limits_{Z\rightarrow\pm\infty}ig\dt_\mu g\equiv i\xi_\pm(x^\mu)\dt_\mu\xi_\pm^{-1}(x^\mu)
\end{equation}
where $\xi_+\in U(N_f)_L$ and $\xi_-\in U(N_f)_R$. Imposing this behavior as a boundary condition, we can express $A^{(g)}_\mu$ as
\begin{equation}
A_\mu^{(g)}=i\xi_+\dt_\mu\xi_+\psi_+ + i\xi_-\dt_\mu\xi_-\psi_- + \sum\limits_{n=1}^\infty B_\mu(x)^{(n)}\psi_n(Z)
\end{equation}
with 
\begin{align}
\psi_\pm(Z)=\frac{1}{2}\left( 1\pm \frac{2}{\pi}\arctan (Z)\right)~.
\end{align}
The nonnormalizable functions $\psi_\pm(Z)$ satisfy the gauge field equations of motion with boundary 
conditions $\psi_\pm (Z\rightarrow \pm\infty) = 1$ and $\psi_\pm (Z\rightarrow \mp\infty) = 0$. 
From $\xi_\pm(x^\mu)$ we can construct the appropriate chiral field $U(x^\mu)=\xi_+^{-1}\xi_-$ transforming 
as $U\rightarrow g_+ U g_-$ under $(g_+,g_-)\in U(N_f)_L\times U(N_f)_R$. 
Following \cite{Sakai:2004cn}, we use the residual gauge invariance to further fix $\xi_+=U(x^\mu)$ and $\xi_-=1$ so that
\begin{equation}\label{eq:Uexpansion}
A_\mu^{(g)}=iU^\dagger\dt_\mu U\psi_+  + \text{normalizable modes}
\end{equation}
We have 5d field strengths
\begin{align}\label{eq:skyrmefieldstr}
F^{(g)}_{\mu\nu}&=-\left[U^{-1}(x^\mu)\partial_\mu U(x^\mu),U^{-1}(x^\mu)\partial_\nu U(x^\mu)\right]\psi_+(\psi_+-1) \\
F^{(g)}_{Z\mu} &= iU^{-1}(x^\mu)\partial_\mu U(x^\mu) \partial_Z\psi_+~.
\end{align}
The DBI action now explicitly produces the 4d Skyrme model
\begin{align}\label{eq:skyrmeaction}
S_{DBI}=\kappa \int d^4x \tr \left\{ A(U^{-1}(x^\mu)\partial_\mu U(x^\mu))^2+
B\left[U^{-1}(x^\mu)\partial_\mu U(x^\mu),U^{-1}(x^\mu)\partial_\nu U(x^\mu)\right]^2\right\}
\end{align}
with parameters $A$ and $B$ defined by the overlap of the non-normalizable modes with warp factors
associated with the background metric:
\begin{align}
A= M_{KK}^2\int dZ \left(1+Z^2\right)(\partial_Z\psi_+)^2=0.318~M_{KK}^2\\
B= \frac{1}{2}\int dZ {(1+Z^2)^{1/3}}\psi_+^2\left(\psi_+-1\right)^2=0.078 ~.
\end{align}
To match the Skyrme Lagrangian with $U(N_f)$ generators 
normalized to $\tr(T^aT^b)=\delta^{ab}/2$, 
\begin{equation}
S_{Skyrme}=\int d^4x \Tr\left( \frac{f_\pi^2}{4} (U^{-1}\dt_\mu U)^2+\frac{1}{32e^2} [U^{-1}\dt_\mu U, U^{-1}\dt_\nu U]^2\right)
\end{equation}
we
identify
\begin{align}
f_\pi^2=4\kappa A\qquad \frac{1}{e^2}=32\kappa B~.
\end{align}
It is easy to show that taking $U(x^\mu)=\exp [-2i\pi(x^\mu)^a T^a/f_\pi]$ and expanding the action to leading order in
the pion field indeed gives a canonical kinetic term for the pion. 
This is the form of the famous Skyrme model \cite{Skyrme:1961vq,Adkins:1983ya}, in which baryons appear as solitonic configurations of $U(x^\mu)$. 
The second term in the Lagrangian (the ``Skyrme term'') stabilizes solitons of finite size. Including the $\omega$ meson
and a gauged WZW term can also be used for this purpose \cite{Adkins:1983nw}.

We are now in a position to fix the remaining free parameters from experimental data.
Following \cite{Sakai:2004cn}, we use the $\rho$ meson mass $m_\rho$ to fix $M_{KK}=949\text{ GeV}$ and the
pion decay constant $f_\pi=93\text{ MeV}$ to fix $\kappa=7.45\times 10^{-3}$ \footnote{The string length $l_s$ remains a free parameter, but none of the quantities relevant to the present analysis depend on it directly.}. It should be noted that this is the crudest way to
set the model parameters, and is intended only to yield a heuristic estimate of the values we can derive from the Sakai-Sugimoto model. More accurate results could be obtained by fitting to multiple real world parameters (such as several more (axial) vector meson masses). An 
analysis of this type is conducted in the ``hard-wall'' model of \cite{Erlich:2005qh}. 

\subsection{Predictions for Free Parameters}
Having detailed the various ingredients of the Sakai-Sugimoto gravity dual, we can now make predictions for three of
the four parameters appearing in our ansatz for small $t$, large $s$ proton-proton scattering: the ratio $-a_0/a'$, the dipole 
mass $M_d$  in the gravitational form factor, and the coupling $\lambda$ between the proton and the $2^{++}$ glueball. 
The first two quantities we take from the existing literature, and calculate the third directly.
\begin{enumerate}
\item Using $M_{KK}=949~\text{ MeV}$ and  (\ref{eq:gluemass}) we estimate $m_g\approx 1.49~\text{GeV}$. This value
is significantly lower than the lattice result,  $m_{g-lattice}=2.40~\text{GeV}$ \cite{Morningstar:1999rf}. In the 
next section, we find that our smaller value of  $m_g^2$ more closely approximates the 
ratio $-a_0/a'$ we find by fitting our model to $pp$ and $p\bar{p}$-scattering data.
\item We model protons as 4d Skyrmions, for which the matrix elements of the energy momentum tensor (EMT)
have been computed explicitly \cite{Cebulla:2007ei}. The form factors can be related to the components of
the static EMT in Breit frame (that is with $k^0=0$):
\begin{equation}
T_{\mu\nu}({\bf r},{\bf s})=\frac{1}{2 p^0}\int \frac{d^3 k}{(2\pi)^3} e^{i{\bf k}\cdot {\bf r}}\langle p',s' | T_{\mu\nu}(0) |p,s \rangle
\end{equation}
where the proton spin polarizations are defined to be equal in the respective protons' rest frames and equal to $s=(0,{\bf s})$. The form factors
are given by 
\begin{eqnarray}
A(t)+B(t)+\frac{2t}{3}(A'(t)+ B'(t))=\int d^3 {\bf r} e^{-i{\bf k}\cdot {\bf r}}\epsilon^{ijk} s_i r_j T_{{\bf r}, {\bf s}} \\
A(t)+\frac{t}{4 m^2}\left[ A(t)+2 B(t)+4 C(t)\right]=\frac{1}{m}\int d^3{\bf r} e^{-i{\bf r}\cdot {\bf k}}T_{00}({\bf r},{\bf s})\\
C'(t)+\frac{4t}{3}C'(t)+\frac{4t^2}{15}C''(t)=-\frac{m}{10}\int d^3 {\bf r} e^{-i{\bf r}\cdot {\bf k}} T_{ij}({\bf r})\left( r^ir^j-\frac{{\bf r}^2}{3}\delta^{ij}\right)
\end{eqnarray}
with $m$ the nucleon mass, and primes denoting differentiation with respect to $t$. 
Evaluating the Skyrme EMT on the hedgehog solution of \cite{Skyrme:1961vq},
$U=\exp [i\tau\cdot\hat{r} F(r)]$ with the radial function $F(r)$ chosen to minimize the soliton mass. \cite{Cebulla:2007ei} determine the
form factors explicitly in the large $N_c$
limit. A dipole form approximates $A(t)$ well for
 up to $|t|< 0.8~\text{GeV}^2$, with dipole mass  $M_{d}=1.17~\text{GeV}$. This value for $M_{d}$ is in good agreement
 with the value obtained by fitting to data, as presented in the next section.
 
 A more rigorous analysis would treat the protons
 as 5d solitons stabilized by vector mesons via the Chern-Simons term. This is beyond the scope of the present work
 and we take the ordinary 4d Skyrme model to be sufficient to provide a heuristic estimate for $M_d$.

 \item We now compute the glueball-proton coupling in the Sakai-Sugimoto model, and find that the glueball indeed couples 
 primarily to the 4d energy-momentum tensor. 
 
 Let us consider graviton couplings in the DBI action, $h_{\mu\nu}$. Keeping only couplings linear in $h$,   
\begin{align}
S_{D8}&=-T_8\int d^9x e^{-\phi} \tr\sqrt{-det(\widetilde{G}_{MN}-h_{MN}+2\pi\alpha'F_{MN})} \nonumber \\
&= -\frac{1}{2}T_8(2\pi\alpha')^2\int d^9x 
e^{-\phi}\sqrt{\widetilde{G}}\widetilde{G}^{\mu\alpha}\widetilde{G}^{\nu\beta}h_{\alpha\beta}(x,z)
\tr\left(\widetilde{G}^{\gamma\delta}F_{\mu\gamma}F_{\nu\delta}+\widetilde{G}^{ZZ}F_{\mu z}F_{\nu z}\right) +\dots \nonumber \\
&=-\kappa \int d^4x  h_{\alpha\beta}(x) \tr\left( A_h(U^{-1}\partial^\alpha U)(U^{-1}\partial^\beta U)+
B_h\left[U^{-1}\partial^\alpha U,U^{-1}\partial_\rho U\right]\left[U^{-1}\partial^\beta U,U^{-1}\partial^\rho U\right]\right)+\dots \nonumber \\
&\equiv -h_{\alpha\beta}C^{\alpha\beta}+\dots \label{eq:gravskyrmecoupling} 
\end{align}
To arrive at the third line, we inserted the expressions for the field strengths in terms of $U(x)$ (eq. \ref{eq:skyrmefieldstr}), and 
$h_{\alpha\beta}(x,Z)$, the pullback of the lightest graviton mode onto the branes.
The
coefficients $A_h$ and $B_h$ are given by the overlap integrals
\begin{align}
A_h=& \frac{18\sqrt{3}\pi M_{KK}}{N_c^{3/2}g_{YM}}\int_{-\infty}^\infty dZ K(Z) \frac{T(Z)}{{\cal I}_T} (\dt_Z\psi_+)^2=13.42\frac{M_{KK}}{g_{YM} N_c^{3/2}} \\
B_h=& \frac{9\sqrt{3}\pi}{N_c^{3/2}g_{YM} M_{KK}}\int_{-\infty}^{\infty}dZ K(Z)^{-1/3} \frac{T(Z)}{{\cal I}_T}\psi_+^2(\psi_+-1)^2=7.64\frac{1}{g_{YM}N_c^{3/2}M_{KK}}~.
\end{align}

Let us compare the tensor $C^{\alpha\beta}$ in (\ref{eq:gravskyrmecoupling}) to the energy momentum tensor (EMT) of the 4d Skyrme model, 
\begin{align}
T_{\alpha\beta}=\kappa \tr\left\{ A (U^{-1}\partial_\alpha U)(U^{-1}\partial_\beta U)
+2B\left[U^{-1}\partial_\alpha U,U^{-1}\partial_\rho U\right] \left[U^{-1}\partial_\beta U,U^{-1}\partial^\rho U\right] \right\}~.
\end{align}
We can extract the coupling constant $\lambda$ as the ratio of the coefficients of either of the two terms in $C_{\alpha\beta}$ and $T_{\alpha\beta}$, or
some linear combination of both.  For example we can consider
 \begin{align}
 C^{\alpha\beta}=\frac{A_h}{A}T^{\alpha\beta}+\kappa\left( B_h-\frac{2BA_h}{A}\right)\left[U^{-1}\partial^\alpha U,U^{-1}\partial_\rho U\right]\left[U^{-1}\partial^\beta U,U^{-1}\partial^\rho U\right]
 \end{align}
 where
 \begin{align}
& \frac{A_h}{A}=42.18 \frac{1}{M_{KK}g_{YM}N_c^{3/2}}= 3.61 ~\text{GeV}^{-1}\\
&\left(B_h-\frac{2BA_h}{A}\right)=0.09~ \text{GeV}^{-1}
 \end{align}
Or alternatively
 \begin{align}
 C^{\alpha\beta}=\frac{B_h}{2B}T^{\alpha\beta}+\kappa\left(A_h-\frac{AB_h}{2B}\right)(U^{-1}\partial^\alpha U)(U^{-1}\partial^\beta U)
 \end{align}
where
\begin{align}
&\frac{B_h}{2B}=48.86 \frac{1}{N_c^{3/2}g_{YM}M_{KK}}=4.19~\text{ GeV}^{-1}\\
&\left(A_h-\frac{AB_h}{2B}\right) =-0.17~ \text{GeV}~. \\
\end{align}
Assuming the relative contributions of the two terms (kinetic and Skyrme) are of the same order, the deviation of $C_{\alpha\beta}$ from 
the EMT amounts to a few percent of the value of $\lambda$. We therefore estimate  $\lambda \sim 3.9\pm 0.3~\text{GeV}^{-1}$, which is within the same order of magnitude as the value derived from the fits discussed in
the next section. It should be kept in mind however that the Skyrme model itself is only accurate to within $\sim 25\%$, and the Skyrme
model is itself only a crude approximation to a soliton that is actually five-dimensional and contains the full five-dimensional gauge fields
on the flavor brane. The value of $\lambda$ derived from 
the Skyrme model may thus deviate significantly from its true value.
\end{enumerate}

\section{Data Fitting and Comparison}

\subsection{Regime of validity}

\begin{figure}
\resizebox{6in}{!}{\includegraphics{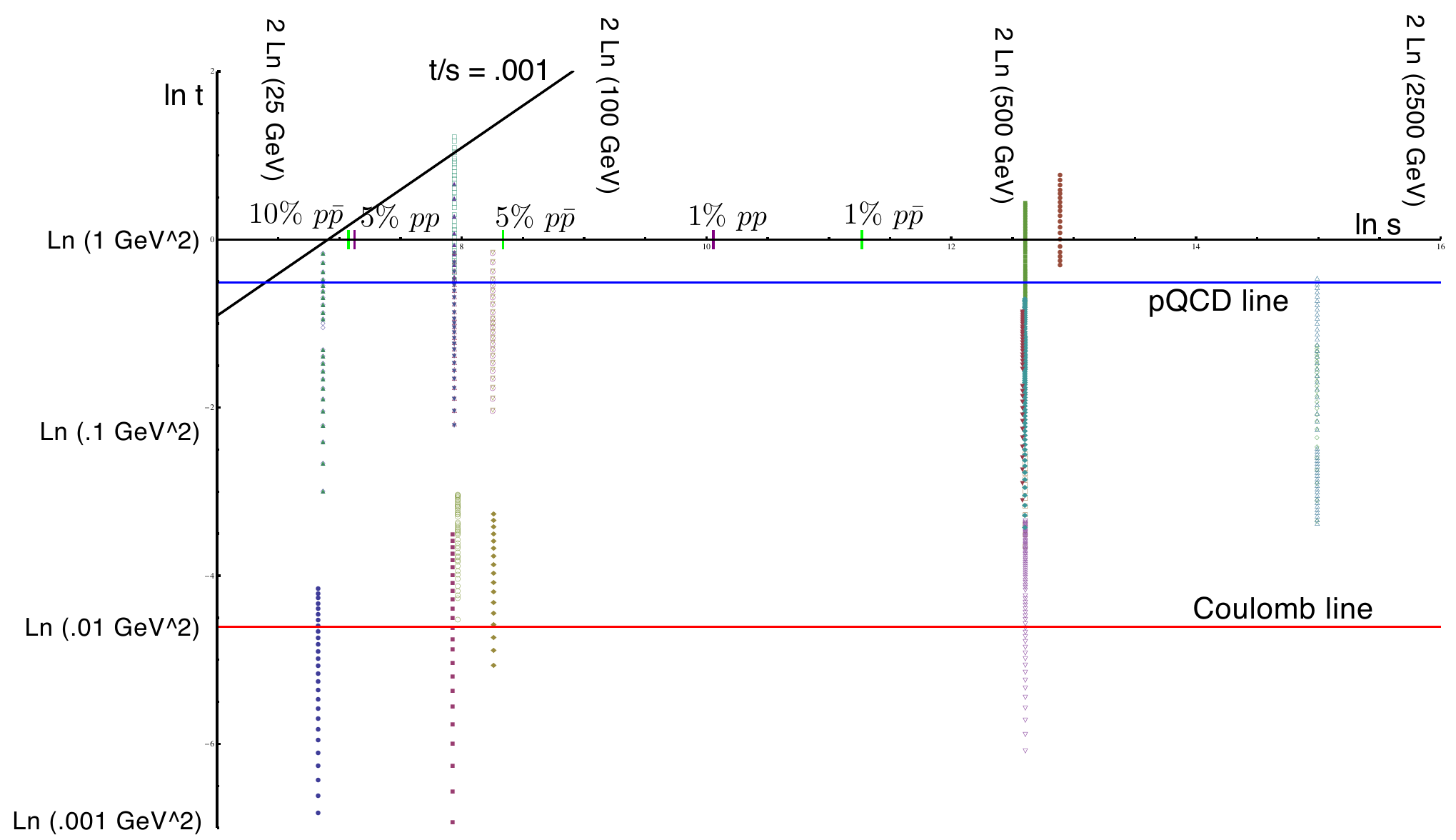}}
\caption{\label{scatterplot} Data from proton-proton and proton-antiproton scattering, with a dot for every data point in $(\log s, \log t)$ space.  A given experiment is generally run at fixed $s$, for a range of values of $t$, so it creates a vertical series of data points.  Below the diagonal line $t/s$ corrections make up less than $0.1 \%$.  The horizontal line near the bottom lies at $|t| =0.01~\text{GeV}^2$, which we are using as a lower cutoff.  Below this line there are significant effects from Coulomb scattering.  The horizontal line near the top is at $|t| = 0.6~\text{GeV}^2$, which we use as an upper cutoff. Along the $\log s$ axis we mark the values at which the Reggeon contribution to proton/proton or proton/anti-proton scattering is $1 \%$, $5 \%$, and $10 \%$.}
\end{figure}

Before using our model to fit experimental scattering data, we briefly discuss the data we use, and limitations in applicability of our model as a function of $s$ and $t$.  
We fit the differential cross section $d\sigma/dt$ for a variety of values of $s$ and $t$ (shown in Fig. \ref{scatterplot}).  
There are many complicating factors which limit the validity of our model, the most obvious being that we have worked in the
strict Regge limit, neglecting corrections suppressed by powers of $t/s$.  For all of the data we use, $t/s<.001$. Other effects, such as Coulomb contributions,
perturbative QCD effects, and the contribution of Reggeon trajectories are not so easily discarded. We discuss each of these in some detail below.

\subsubsection{Coulomb Contributions}

For very small values of $|t|$ (regardless of $s$) the Coulomb interaction makes a significant contribution to the amplitude.  
This contribution will be largest at $|t| \simeq 0.002 \ \text{GeV}^2$ and is
negligible (for our purposes) by $t \simeq 0.01 \ \text{GeV}^2$ \cite{Amos:1985wx, Bernard:1987vq}.  We will use $|t| =0 .01 \ \text{GeV}^2$ 
as a lower cutoff in $t$.

\subsubsection{Lower Regge trajectories}

For large values of $s$, the total cross sections converge because the exchange of a Pomeron does not distinguish between particles and antiparticles.  
For small values of $s$, however, there are contributions to both proton-proton and proton-antiproton scattering from other Regge trajectories.  
In particular, the lower Regge trajectory is actually a pair of exchange degenerate trajectories, one consisting of even spin particles and the other of odd spin particles.  For proton-antiproton scattering the Reggeon contribution is larger because these two trajectories add, whereas for proton-proton scattering 
they work to cancel each other out.  
We can estimate how large the contributions from the next Regge trajectory will be by looking at the total cross sections for proton-proton and proton-antiproton scattering.  

\begin{figure}
\resizebox{6in}{!}{\includegraphics{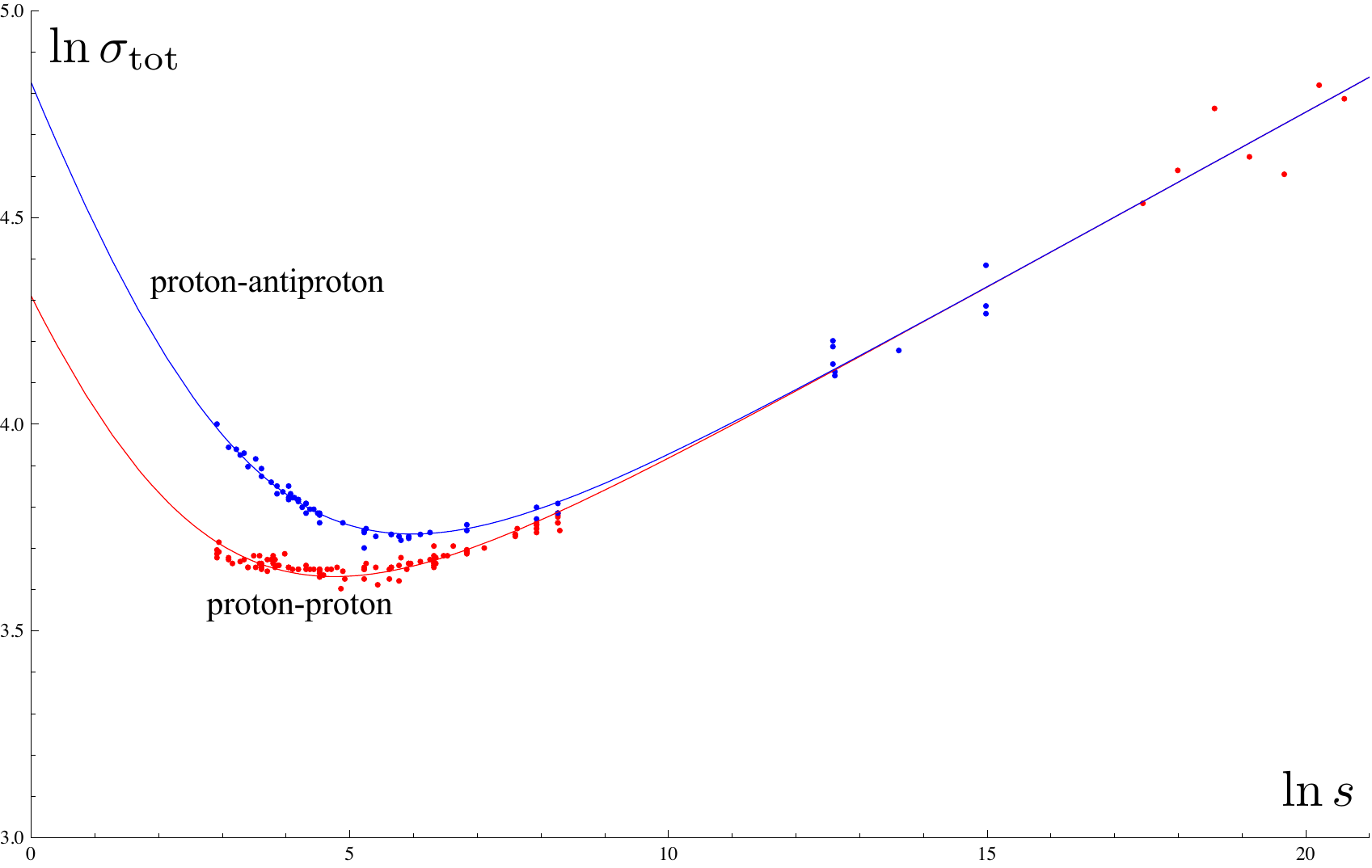}}
\caption{\label{total} Log-log plot of total cross sections (in mb) for proton-proton and proton-antiproton scattering as a function of $s$ (in $\text{GeV}^2$). Note that the cross sections converge
as $s$ grows large.}
\end{figure}

We fit the total cross sections with the functions
\beq
\sigma_{\mathrm{tot}}(pp \rightarrow pp) = Ps^{p} + Qs^{q}
\eeq
\beq
\sigma_{\mathrm{tot}}(p\bar{p} \rightarrow p\bar{p}) = Ps^{p} + Rs^{q}
\eeq
and find best fit values
\beq
p = .08, \hspace{.5in} q = -.46, \hspace{.5in} P = 21.3, \hspace{.5in} Q = 53.2, \hspace{.5in} R = 103.6.
\eeq

We can use the optical theorem to relate the total cross section to the differential cross section at $t = 0$, and use this in turn to estimate the size of the first contribution from the lower trajectory.  
\beq
\frac{d\sigma}{dt}(pp (p\bar{p})) \sim P^2 s^{2p}\left(1 + \frac{2Q(R)}{P}s^{q - p} + \cdots\right).
\eeq
Based on this functional form,  the magnitude of Reggeon contamination in proton/anti-proton scattering at $\sqrt{s} = 31$ GeV is about $22\%$, while at $\sqrt{s} = 1800$ GeV 
it is about $0.3\%$.  In FIG. \ref{scatterplot}, the marks on the $\log s$ axis show where Reggeon contamination to proton-antiproton and proton-proton scattering is 
$1\%$, $5\%$ and $10\%$.  For the lower center-of-mass energy data, the effect is fairly large.  We could account for this by adding to our model a term corresponding to Reggeon exchange.  For the present treatment, however, we simply add the amount of Reggeon contamination for a given value of $s$ to the experimental 
error associated with each data point, thus weighting our fit towards the higher energy data.

\subsubsection{Perturbative QCD and the hard Pomeron}

\begin{figure}
\resizebox{4in}{!}{\includegraphics{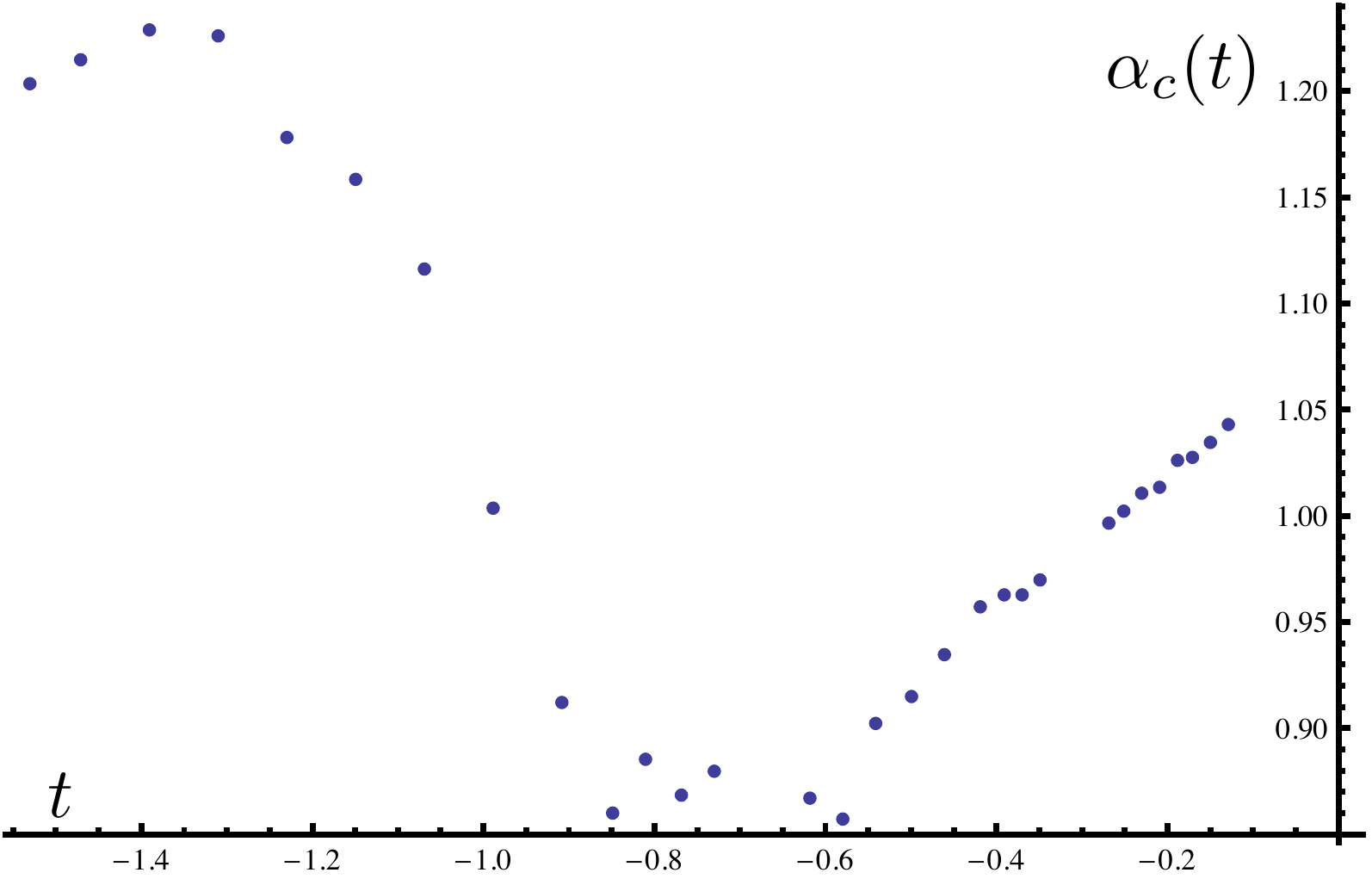}}
\caption{\label{traj} The function $\alpha_c(t)$.  In the region $0 \le |t| \le 0.6 \ \text{GeV}^2$ the trajectory matches that of the soft Pomeron.  Outside this region we assume that perturbative QCD effects make our treatment inapplicable.}
\end{figure}

For sufficiently small values of $|t|$,  it is reasonable to think of the Pomeron as a trajectory of confined states (glueballs). This is the ``soft Pomeron'' on
which our model is based.  As $|t|$ increases, however, we eventually enter the regime of perturbative QCD, outside our model's regime of validity. 
It is not clear on theoretical grounds exactly where this transition occurs; we instead attempt to determine its location empirically, by
examining the data.  
Let us assume that the Pomeron contribution to the differential cross section is of the form
\beq
\label{basicform}
\frac{d\sigma}{dt} = F(t)s^{2\alpha_c(t) - 2}
\eeq
for all values of $t$,
where $F(t)$ is some unknown function.  Differential cross section data is  typically 
analyzed for a range of $t$ values at a fixed value of $s$.  Suppose instead we consider a fixed value of $t$ for a range of values of $s$.  The trajectory $\alpha_c(t)$ is therefore the slope of $\log (d\sigma/dt)$ plotted as a function of  $log(s)$:
\beq
\log \frac{d\sigma}{dt} = \log F(t) + (2\alpha_c(t) - 2)\log s.
\eeq
By fitting for the slope of this line at a range of values of $t$, we can get a reasonable picture of the function $\alpha_c(t)$.  Referring again to FIG. \ref{scatterplot}, we can see that taking sets of the data at fixed values of $|t|$ is difficult with the extant data, and generally these sets will only have between 3 and 5 data points each.  This grouping of the data would not yield reliable statistics, so we do not use it to fit $\alpha_c(t)$ directly, but consider it a reasonable estimate of where the transition occurs between the regime where the soft Pomeron accurately characterizes the exchanged degrees of freedom, and the regime where it does not. A graph of $\alpha_c(t)$ as a function of $|t|$ is shown in FIG. \ref{traj}.  We can see that for $0 \le |t| \le 0.6 \ \text{GeV}^2$ the trajectory matches what we expect for the soft Pomeron.  There is a clear transition at $|t| = 0.6 \ \text{GeV}^2$, where the slope suddenly
becomes much less steep; above $|t|>0.9\ \text{GeV}^2$ the behaviour is clearly nonlinear. We therefore attach no particular significance to the shape of the plot in this region, choosing instead to impose an upper bound of $0.6 \ \text{GeV}^2$ on the values of $|t|$ we consider in our fits.

\subsection{Comparison to a ``Photo-Pomeron'' model}

In order to provide a frame of reference for the success of the model described above compared to the existing literature, we briefly review
a commonly-used model for single Pomeron exchange due to Donnachie, Jaroszkiewicz, Landshoff, Polkinghorne and 
others (\cite{Donnachie:1983hf}, \cite{Landshoff:1974db}, \cite{Jaroszkiewicz:1974ep}).
This is only one of many possible models of various degrees of complication, but serves as an illustrative
example because it has the same structure as our model, with the difference that it relies on the electromagnetic
proton form factor rather than the gravitational form factor. Other models based on different assumptions include
the impact picture model of \cite{Bourrely:1984gi}, the multi-component model of the Durham group (see 
\cite{Ryskin:2009tj,Ryskin:2009tk} for a recent discussion) and the eikonal model of Block, Halzen and collaborators
(see e.g. \cite{Block:1998hu} and references therein). 

The electromagnetic-type Pomeron coupling (reviewed in \cite{DLbook}) draws inspiration from the additive quark rule: the (experimental) fact that
the ratios of total cross section equal the ratios of the numbers of valence quarks
present in the scattered hadrons.
Positing 
that the Pomeron couples to 
constituent quarks individually as $\gamma^\mu$ reproduces this observation.
The form of the Pomeron-quark coupling is then assumed to be identical to the photon coupling
except that the Pomeron is $C=+1$ rather than
$C=-1$. The form factor $F_1(t)$ involved in the exchange is assumed to be identical to 
the electromagnetic form factor. 
For $|t|<1~\text{GeV}^2$, 
a dipole approximation to $F_1(t)$ (from electron scattering data)   gives
 \begin{equation}\label{eq:electromagformfactor}
 F_1(t)=\frac{4m_p^2-2.79t}{4m_p^2-t}\frac{1}{(1-t/0.71)^2}~.
 \end{equation}
Using this single-Pomeron exchange model, the unpolarized $pp$ (or $p\pB$)
 cross section becomes 
 \begin{equation}\label{eq:DL}
\frac{d\sigma}{dt}=\frac{(3\beta_P F_1(t))^4}{4\pi}\left(\frac{s}{s_0}\right)^{2\alpha_P(t)-2}~,
\end{equation}
with $s_0\approx 1 \ \text{GeV}^2$ the characteristic scale of the problem.
The key difference between this model and ours lies in the form factor: gravitational in our case,
electromagnetic in the case of the ``photo-Pomeron.''

\subsection{Fits to Scattering Data}

\begin{figure}
\resizebox{6in}{!}{\includegraphics{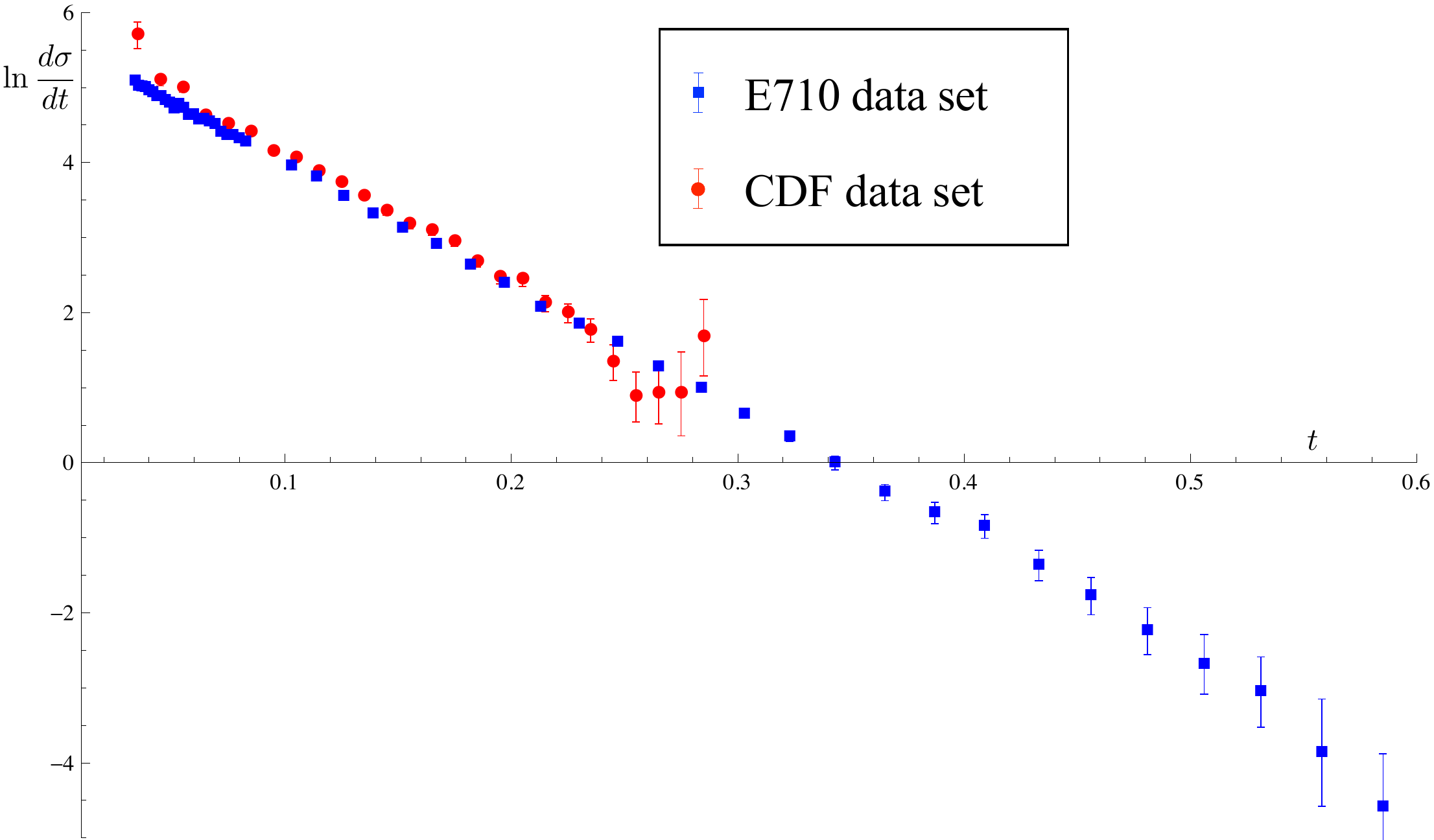}}
\caption{\label{CDFvsE710} A log-linear plot of the differential cross section at $\sqrt{s}=1800\text{ GeV}$ from the CDF and E710 collaborations,  as a function of $t$ (in $\text{GeV}^2$) .}
\end{figure}

We perform standard  least squares fits \footnote{All fits are completed using Python MINUIT.} 
to the differential cross section data for $pp$ and $p\bar{p}$ scattering using the form (\ref{eq:dhm_diffxsec}), allowing
the parameters $\alpha_0$, $\alpha'$, $M_d$, and $\lambda$ to vary. We also perform an identical fit for the photo-Pomeron model,
allowing $\beta_P$, $\alpha_0$ and $\alpha'$ to vary. 
The data is taken from the Durham HEP database (http://durpdg.dur.ac.uk) with $0.01<t<0.6 \  \text{GeV}^2$ and $30.4<\sqrt{s} <1800\text{ GeV}$.
As our model does not take into account the effects of Reggeon exchange, we simply estimate the contribution of
Reggeons to $d\sigma/dt$ at particular values of $\sqrt{s}$, and add the result 
in quadrature to the experimental errors. There is significant disagreement between two data sets at $\sqrt{s}=1800\text{ GeV}$
produced by the CDF and E710 experiments, as shown in FIG. \ref{CDFvsE710} . Rather than choosing one or the other set explicitly, we perform all
fits using both, just CDF, and just E710, with results as displayed in Table \ref{table:DHMvsDLfits}. 

\begin{figure}[!h]
\resizebox{6.5in}{!}{\includegraphics{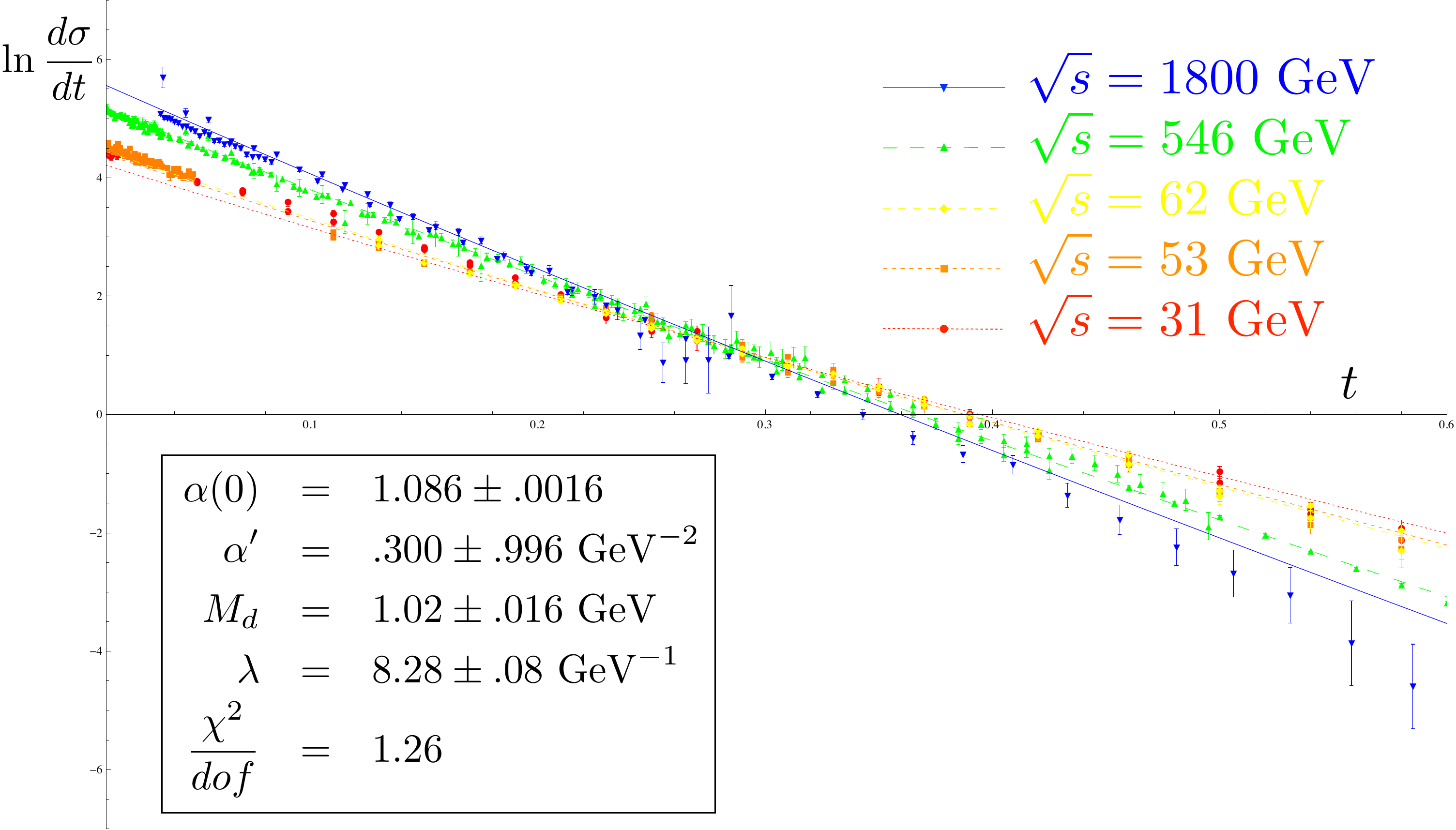}}
\caption{\label{bestfit} A log-linear plot of the best fit to the scattering data using just the CDF data at $\sqrt{s} = 1800$ GeV.  The differential cross 
section is in $\text{mb}/\text{GeV}^2$ and $t$ is in $\text{GeV}^2$.  Note that at lower center of mass energies the 
fit is less successful, most likely due to a greater contribution from Reggeon exchange.}
\end{figure}

\begin{table}
$$
\begin{array}{|rcl|rcl|rcl|}
\multicolumn{9}{c}{\mathrm{Gravitationally\ coupled\ Pomeron}} \\
\hline
\multicolumn{3}{|c|}{\mathrm{both \ data \ sets}} & \multicolumn{3}{c}{\mathrm{just \ E710}} & \multicolumn{3}{|c|}{\mathrm{just \ CDF}} \\
\hline
\alpha_c(0) & = & 1.076 \pm .0016 & \alpha_c(0) & = & 1.074 \pm .0016 & \alpha_c(0) & = & 1.086 \pm .0016 \\
\alpha_c' & = & .290 \pm .006 \ \mathrm{GeV}^{-2} & \alpha_c' & = & .286 \pm .006 \ \mathrm{GeV}^{-2} & \alpha_c' & = & .300 \pm .006 \ \mathrm{GeV}^{-2} \\
M_d & = & .983 \pm .016 \mathrm{GeV} & M_d & = & .970 \pm .016 \ \mathrm{GeV} & M_d & = & 1.02 \pm .016 \ \mathrm{GeV} \\
\lambda & = & 8.56 \pm .08 \ \mathrm{GeV}^{-1} & \lambda & = & 8.62 \pm .08 \ \mathrm{GeV}^{-1} & \lambda & = & 8.28 \pm .08 \ \mathrm{GeV}^{-1} \\
\frac{\chi^2}{d.o.f} & = & 1.65 & \frac{\chi^2}{d.o.f} & = & 1.41 & \frac{\chi^2}{d.o.f.} & = & 1.26 \\
\hline
\multicolumn{9}{c}{} \\
\multicolumn{9}{c}{\mathrm{Electromagnetically\ coupled\ Pomeron}} \\
\hline
\multicolumn{3}{|c|}{\mathrm{both \ data \ sets}} & \multicolumn{3}{c}{\mathrm{just \ E710}} & \multicolumn{3}{|c|}{\mathrm{just \ CDF}} \\
\hline
\alpha_c(0) & = & 1.076 \pm .0013 & \alpha_c(0) & = & 1.075 \pm .0013 & \alpha_c(0) & = & 1.082 \pm .0018 \\
\alpha_c' & = & .289 \pm .003 \ \mathrm{GeV}^{-2} & \alpha_c' & = & .289 \pm .003 \ \mathrm{GeV}^{-2} & \alpha_c' & = & .289 \pm .003 \ \mathrm{GeV}^{-2} \\
\beta & = & 1.858 \pm .016 \ \mathrm{GeV}^{-1} & \beta & = & 1.877 \pm .016 \ \mathrm{GeV}^{-1} & \beta & = & 1.801 \pm .020 \ \mathrm{GeV}^{-1} \\
\frac{\chi^2}{d.o.f} & = & 1.97 & \frac{\chi^2}{d.o.f} & = & 1.66 & \frac{\chi^2}{d.o.f.} & = & 1.79 \\
\hline
\end{array}
$$
\caption{\label{table:DHMvsDLfits} Fits to differential cross section data including both
E710 and CDF $\sqrt{s}=1800$ data sets, just E710, or just CDF.}
\end{table}
Our model clearly 
produces the smallest value of $\chi^2 / d.o.f.$ when only the CDF data set is included, but fares better than 
the photo-Pomeron model in all cases.  FIG. \ref{bestfit} shows fits of our model to the differential cross section data.

We can also compare our predictions for the total cross section and $\rho=\text{Re}{\cal A}(t=0)/\text{Im}{\cal A}(t=0)$ 
 (based on the best-fit parameters determined from the 
differential cross section) to data. Applying the optical theorem to eq. (\ref{eq:dhm_amp}) yields the total cross section
\begin{equation}
\sigma_{tot}=\frac{\pi\lambda^2\Gamma [-\chi]}{\Gamma \left[\frac{\alpha_c(0)}{2}\right]\Gamma \left[\frac{\alpha_c(0)}{2}-1-\chi \right]} 
\left(\frac{\alpha_c' s}{2} \right)^{\alpha_c(0)-1} \equiv C s^{\alpha_0-1}
\end{equation}
where in terms of the best fit parameters (using only CDF data) we find $C=21.325$ and $\alpha_0-1=0.085$. Performing an explicit
$\chi^2$ fit to total cross section data, we find $C_{fit}=21.097$ and $(\alpha_c(0)-1)_{fit}=0.086$, in excellent agreement with the computation.

\begin{figure}
\resizebox{4in}{!}{\includegraphics{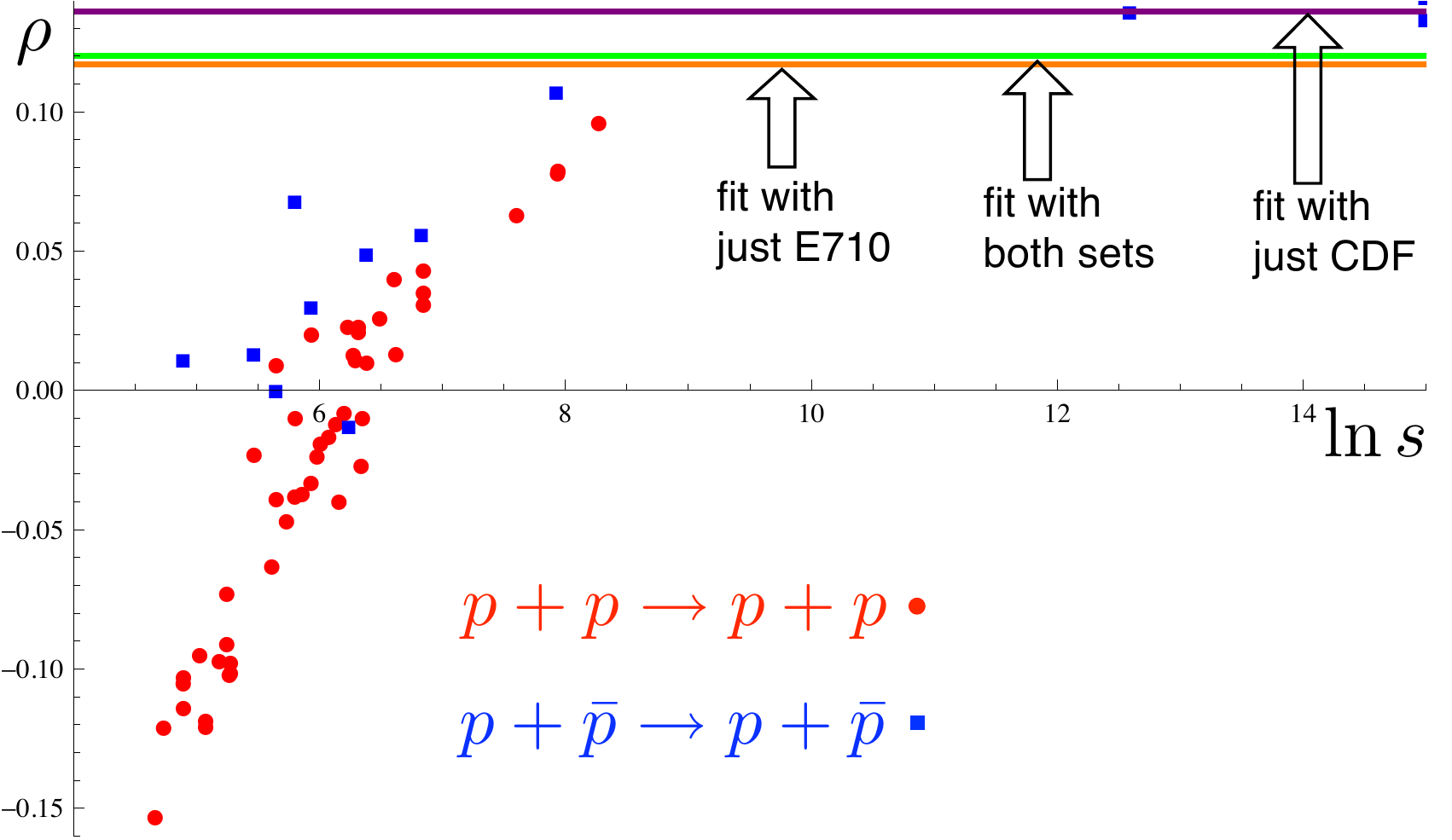}}
\caption{\label{rhograph} Data for $\rho$ as a function of $\log s$ compared to the predicted value.}
\end{figure}

Because we neglect Reggeons, (\ref{eq:dhm_amp}) predicts a constant value for $\rho$ as $\rho=-\cot \left( \frac{\pi\alpha_0}{2}-\pi\right)=0.136$ (where
again we use the values of the best-fit parameters from the differential cross section).
This value agrees well with the data at large $\sqrt{s}$ (see Fig. \ref{rhograph}), where the Reggeon contribution is minimal.

We have shown that our mechanism for Pomeron exchange in $pp$ and $p\bar{p}$ scattering fits experimental data quite well. 
The best fit parameters from fitting the data also compare favorably with our estimates for $a_c(0)/a_c'$, $\lambda$, and
$M_d$ computed in holographic QCD. The Sakai-Sugimoto model predicts the mass of the lightest spin 2 glueball  $m_g=1.485\ \text{GeV}$,
while the value produced by fitting the form of the differential cross section to scattering data yields $m_{g-fit}=1.745\pm0.035 \ \text{GeV}$, 
is within $15\%$ of the computed mass, though not with the statistical error bars determined by the fit. 
The gravitational dipole mass $M_d$ computed in the Skyrme model has value $M_d=1.17 \ \text{GeV}$,
which deviates from the fitted value $M_{d-fit}=1.02\pm 0.016$ by about $15\%$ as well. As the Skyrme model predicts masses only to an
accuracy of about $20\%$, the fitted value lies within the expected uncertainty of the computed dipole mass. Finally, the coupling constant 
computed from holography to be $\lambda\approx 3.90\pm 0.3 \ \text{GeV}^{-1}$ is within the same order of magnitude as the 
best fit value of $8.28\pm 0.08~ \text{GeV}^{-1}$. We stress again that in this case the $\lambda$ computed in Sakai-Sugimoto is only a crude estimate, and at this level we should be content with order-of-magnitude agreement.

We should note that the values of $\chi^2$ per d.o.f. we obtain are not as good as those of fits cited in the current PDG which typically fit to a leading $\log^2s$ behavior and use more sophisticated filtering of the data than we have done \cite{Cudellpdg,Igipdg,Blockpdg}.

\begin{figure}
\resizebox{4in}{!}{\includegraphics{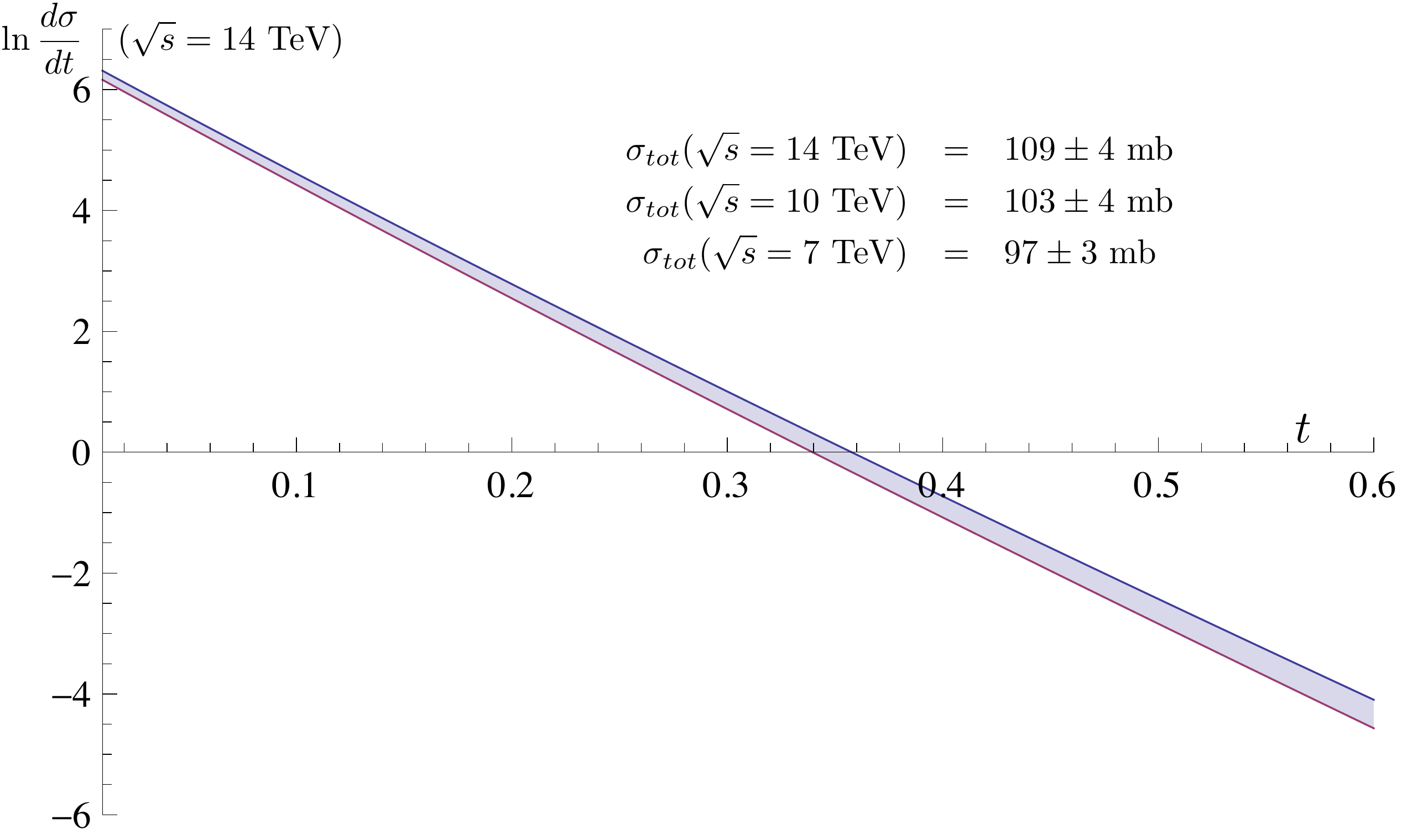}}
\caption{\label{LHCpredict} A log-linear plot of the predicted LHC differential cross section (in $\text{mb}/ \text{GeV}^2$) as a function of $t$ (in $\text{GeV}^2$) at $\sqrt{s}=14~\text{TeV}$. We have used the best fit
parameters from the data set including only the CDF $\sqrt{s}=1800~\text{GeV}$ data. The range shown is
generated by the errors in the fit parameters. }
\end{figure}

Finally, we can use our model for $pp$ scattering to make a prediction for the differential and total cross sections to be observed at the
LHC (at $\sqrt{s}=14~ \text{TeV}$). The differential cross section $d\sigma/dt$ is plotted in Fig. \ref{LHCpredict} as a function of the momentum transfer, $t$. We predict a total cross section of $\sigma_{tot}=109\pm4\text{mb}$.

\section{Conclusions and future directions}

We have used AdS/QCD to construct a general model for Pomeron exchange in Regge-limit $p-p$ and $\bar p-p$ scattering. 
In order to do so, we assumed that string scattering amplitudes have the same structure in weakly curved space as they do in flat space, but that
the \textit{values} of the Regge trajectory parameters and the masses of excitations are modified. 
This implies that the curved space Regge regime amplitude factorizes into a piece that characterizes the interaction of the scattered particles with the exchanged trajectory, and a piece that corresponds simply to the exchange of a closed string (the Reggeized
propagator). Furthermore, the coupling of the external states to the exchanged trajectory is described entirely by their coupling to the lowest mode (the $2^{++}$ glueball). 

Using  these principles, and generic properties of QCD duals, 
we were able to identify the form factor $A(t)$ at the Pomeron-proton vertex
as coming from the matrix element of the energy-momentum tensor, with the strength of the coupling given by an overlap of
graviton and pion wave functions. In a particular hQCD dual, the Sakai-Sugimoto model, we computed the coupling directly and showed that it agrees with one determined by fits to experimental data.
We also developed a method for extending the amplitude for
exchanging the $2^{++}$ glueball to include the exchange of the entire glueball trajectory. 

Though our treatment offers several advantages compared to previous approaches to this problem, there are some ways in which our methodology could be improved or extended.  At the level of numerical analysis, our errors did not take into account systematic errors that might drive an entire data set at a particular $\sqrt{s}$ either up or down.  More sophisticated fitting techniques would also filter
outliers out of the data, which could significantly improve the $\chi^2$ we find.

On a theoretical level, one could certainly extend the regime of validity of our model to lower values of $\sqrt{s}$ by modeling Reggeon (open string) 
exchange as well as Pomeron exchange. Our treatment of the proton in the dual model was also rather
simplistic. Computing the proton form factors and proton-proton-glueball coupling using the duals of baryons as 5d
instantons in the Sakai-Sugimoto model \cite{Hata:2007mb,Hashimoto:2008zw} rather than via the simple 4d Skyrme model, would yield a 
more accurate holographic picture; one should
also take into account that the 5d solitons may be
stabilized using vector meson modes. As mentioned above, \cite{Abidin:2009hr} presented a treatment of protons as fermionic fields in bottom-up holographic
models, which included results for the electromagnetic and gravitational form factors. 
In a recent note, \cite{Domokos:2010ma}, we computed the value of $\lambda$ using a similar treatment in the Sakai-Sugimoto
model (as described in \cite{Hata:2007mb,Hashimoto:2008zw}). This result, of $\lambda_{fermi}\approx \ 6.38 \ \text{GeV}^{-1}$ is in
much better agreement with the fit value of $\lambda_{fit}\approx \ 8.28 \ \text{GeV}^{-1}$ cited above.

We should also remind the reader that we used the four-point sphere amplitude to model the ``Reggeized propagator.''   In the Regge limit this amplitude
indeed consists of the $t$-channel exchange of a closed string, to which the Pomeron is dual.  However, it is not clear that the incoming
and outgoing particles (the protons) are themselves dual to closed strings. As they may be considered as either to solitonic configurations of open string modes
(pions) or as wrapped D4-branes, we might need to consider a more complicated amplitude to accurately 
reflect the structure of the scattering process.  This is a difficult problem, but additional insight might be gained from 
 $\pi-\pi$ scattering, where the string dual should be an annulus amplitude.

Finally, there is the difficult issue of corrections for large $s$. The Froissart bound indicates that at 
some high $s$, the behavior $s^{\alpha(0 - 1}$ for the total cross section must be replaced by a function that grows at most like $\log^2 s$.  
There are two possible sources for large $s$ corrections to our model: string loop corrections, a.k.a. Regge cuts or multiple Pomeron exchange, and corrections to the string amplitude from the curvature of the AdS space.  It is certainly possible that such effects already play a role at energies we consider. 
More thoroughly examining the effects of spacetime curvature in particular would improve the accuracy of our predictions, and would hopefully serve as 
evidence for the existence of a curved-space string dual to QCD.



%
%


\begin{acknowledgements}
This work was supported in part by
NSF Grants PHY-00506630 and 0529954 and DOE Grant 580093. We thank Jon Rosner for helpful conversations. SD and NM would like to thank Alison Brizius for numerical assistance. JH, NM, and SD acknowledge the
Galileo Galilei Institute for Theoretical Physics, the Kavli Institute for Theoretical Physics, and Imperial College London, respectively, for hospitality during the course of this work.

\end{acknowledgements}




\begin{references}


\bibitem{Manohar}
  A.~V.~Manohar,
  ``Large N QCD,''
  arXiv:hep-ph/9802419.
  \bibitem{Coleman}
  S.~R.~Coleman,
  ``1/N,'' in {\it Aspects of Symmetry: Selected Erice Lectures}, Cambridge University Press, 1985.
\bibitem{Collinsbook}
 P.~D.~B.~Collins,
 {\it An Introduction To Regge Theory And High-Energy Physics,}
Cambridge 1977.
\bibitem{Veneziano}
  G.~Veneziano,
  ``Construction of a crossing - symmetric, Regge behaved amplitude for
  linearly rising trajectories,''
  Nuovo Cim.\  A {\bf 57}, 190 (1968).
 \bibitem{JoeI}
  J.~Polchinski,
  {\it String theory. Vol. 1: An introduction to the bosonic string,}
Cambridge University Press (1998).
\bibitem{lovelace}
C.~Lovelace,
``A novel application of Regge trajectories,''
Phys.\ Lett.\  B {\bf 28}, 264 (1968).
\bibitem{shapiro}
J.~A.~Shapiro,
``Narrow-resonance model with Regge behavior for pi pi scattering,''
Phys.\ Rev.\  {\bf 179}, 1345 (1969).
\bibitem{Chew}
G.~F.~Chew and S.~C.~Frautschi,
``Principle of equivalence for all strongly interacting particles within the
S matrix framework,''
Phys.\ Rev.\ Lett.\  {\bf 7}, 394 (1961).
\bibitem{Gribov}
V.~N.~Gribov, JETP, {\bf 14}, 478 (1961).
%
 \bibitem{CGM}
 P.~D.~B.~Collins, F.~D.~Gault and A.~D.~Martin,
 ``Proton Proton Scattering and the Pomeron,''
Nucl.\ Phys.\  B {\bf 80}, 135 (1974).
%
\bibitem{dl}
  A.~Donnachie and P.~V.~Landshoff,
  ``Total cross-sections,''
  Phys.\ Lett.\  B {\bf 296}, 227 (1992)
  %
 %
\bibitem{Cudell}
J.~R.~Cudell, V.~Ezhela, K.~Kang, S.~Lugovsky and N.~Tkachenko,
``High-energy forward scattering and the Pomeron: Simple pole versus
unitarized models,''
Phys.\ Rev.\  D {\bf 61}, 034019 (2000)
[Erratum-ibid.\  D {\bf 63}, 059901 (2001)]
[arXiv:hep-ph/9908218].
    \bibitem{Wittenqcd}
  E.~Witten,
  ``Anti-de Sitter space, thermal phase transition, and confinement in  gauge theories,''
  Adv.\ Theor.\ Math.\ Phys.\  {\bf 2}, 505 (1998)
  [arXiv:hep-th/9803131].
  %
\bibitem{Sakai:2004cn}
  T.~Sakai and S.~Sugimoto,
  ``Low energy hadron physics in holographic QCD,''
  Prog.\ Theor.\ Phys.\  {\bf 113}, 843 (2005)
  [arXiv:hep-th/0412141].
%
\bibitem{Erlich:2005qh}
  J.~Erlich, E.~Katz, D.~T.~Son and M.~A.~Stephanov,
  ``QCD and a holographic model of hadrons,''
  Phys.\ Rev.\ Lett.\  {\bf 95}, 261602 (2005)
  [arXiv:hep-ph/0501128].
\bibitem{DaRold:2005zs}
L.~Da Rold and A.~Pomarol,
``Chiral symmetry breaking from five dimensional spaces,''
Nucl.\ Phys.\ B {\bf 721}, 79 (2005);
\bibitem{Maldacena:1997re}
  J.~M.~Maldacena,
  ``The large N limit of superconformal field theories and supergravity,''
  Adv.\ Theor.\ Math.\ Phys.\  {\bf 2}, 231 (1998)
  [Int.\ J.\ Theor.\ Phys.\  {\bf 38}, 1113 (1999)];
\bibitem{Gubser}
  S.~S.~Gubser, I.~R.~Klebanov and A.~M.~Polyakov,
  ``Gauge theory correlators from non-critical string theory,''
  Phys.\ Lett.\ B {\bf 428}, 105 (1998);
\bibitem{Witten}
  E.~Witten,
  ``Anti-de Sitter space and holography,''
  Adv.\ Theor.\ Math.\ Phys.\  {\bf 2}, 253 (1998).
%
\bibitem{bpst}
R.~C.~Brower, J.~Polchinski, M.~J.~Strassler and C.~I.~Tan,
``The Pomeron and Gauge/String Duality,''
JHEP {\bf 0712}, 005 (2007)
[arXiv:hep-th/0603115].
\bibitem{glueballs}
  C.~Csaki, H.~Ooguri, Y.~Oz and J.~Terning,
  ``Glueball mass spectrum from supergravity,''
  JHEP {\bf 9901}, 017 (1999)
  [arXiv:hep-th/9806021].
  R.~de Mello Koch, A.~Jevicki, M.~Mihailescu and J.~P.~Nunes,
  ``Evaluation Of Glueball Masses From Supergravity,''
  Phys.\ Rev.\  D {\bf 58}, 105009 (1998)
  [arXiv:hep-th/9806125].
  R.~C.~Brower, S.~D.~Mathur and C.~I.~Tan,
  ``Glueball Spectrum for QCD from AdS Supergravity Duality,''
  Nucl.\ Phys.\  B {\bf 587}, 249 (2000)
  [arXiv:hep-th/0003115].

\bibitem{Meyer:2004jc}
H.~B.~Meyer and M.~J.~Teper,
``Glueball Regge trajectories and the Pomeron: A lattice study,''
Phys.\ Lett.\  B {\bf 605}, 344 (2005)
[arXiv:hep-ph/0409183].
\bibitem{bfkl}
For an introduction see
J. R. Forshaw and D. A. Ross, {\it Quantum Chromodynamics and the Pomeron}, Cambridge University Press (1997).
%
\bibitem{Janik}
  R.~A.~Janik,
  ``String fluctuations, AdS/CFT and the soft pomeron intercept,''
  Phys.\ Lett.\  B {\bf 500}, 118 (2001)
  [arXiv:hep-th/0010069];   
 %
\bibitem{Polyakov:1997tj}
  A.~M.~Polyakov,
  ``String theory and quark confinement,''
  Nucl.\ Phys.\ Proc.\ Suppl.\  {\bf 68}, 1 (1998)
  [arXiv:hep-th/9711002].
  %
\bibitem{Andreev:2004jm}
  O.~Andreev,
  ``More comments on the high-energy behavior of string scattering  amplitudes
  in warped spacetimes,''
  Phys.\ Rev.\  D {\bf 71}, 066006 (2005)
  [arXiv:hep-th/0408158].
  %
  %
\bibitem{Andreev:2004sy}
  O.~Andreev and W.~Siegel,
  ``Quantized tension: Stringy amplitudes with Regge poles and parton
  behavior,''
  Phys.\ Rev.\  D {\bf 71}, 086001 (2005)
  [arXiv:hep-th/0410131].
%
 \bibitem{Wittenbaryon}
  E.~Witten,
  ``Baryons In The 1/N Expansion,''
  Nucl.\ Phys.\  B {\bf 160}, 57 (1979).
\bibitem{freundtmd}
P.~G.~O.~Freund, Phys.\ Lett.\ {\bf 2},136 (1962).
\bibitem{tmdII}
P.~G.~O.~Freund, Phys.\ Rev.\ Lett.\ {\bf 16}, 291 (1966).
\bibitem{Pagels}
  H.~Pagels,
  ``Energy-Momentum Structure Form Factors of Particles,''
  Phys.\ Rev.\  {\bf 144}, 1250 (1966).
 \bibitem{Brodsky:2008pf}
  S.~J.~Brodsky and G.~F.~de Teramond,
  ``Light-Front Dynamics and AdS/QCD Correspondence: Gravitational Form Factors of Composite Hadrons,''
  Phys.\ Rev.\  D {\bf 78}, 025032 (2008)
  [arXiv:0804.0452 [hep-ph]].
  %
\bibitem{Ji:1996nm}
  X.~D.~Ji,
  ``Deeply-virtual Compton scattering,''
  Phys.\ Rev.\  D {\bf 55}, 7114 (1997)
  [arXiv:hep-ph/9609381].
\bibitem{Wittenb}
  E.~Witten,
  ``Baryons and branes in anti de Sitter space,''
  JHEP {\bf 9807}, 006 (1998)
  [arXiv:hep-th/9805112].
\bibitem{Abidin:2009hr}
  Z.~Abidin and C.~E.~Carlson,
  ``Nucleon electromagnetic and gravitational form factors from holography,''
  Phys.\ Rev.\  D {\bf 79}, 115003 (2009)
  [arXiv:0903.4818 [hep-ph]].

\bibitem{Hong}
  S.~Hong, S.~Yoon and M.~J.~Strassler,
  ``On the couplings of vector mesons in AdS/QCD,''
  JHEP {\bf 0604}, 003 (2006)
  [arXiv:hep-th/0409118].
\bibitem{Grigoryan}
  H.~R.~Grigoryan and A.~V.~Radyushkin,
  ``Form Factors and Wave Functions of Vector Mesons in Holographic QCD,''
  Phys.\ Lett.\  B {\bf 650}, 421 (2007)
  [arXiv:hep-ph/0703069].
\bibitem{Cebulla:2007ei}
  C.~Cebulla, K.~Goeke, J.~Ossmann and P.~Schweitzer,
  ``The nucleon form-factors of the energy momentum tensor in the Skyrme
  model,''
  Nucl.\ Phys.\  A {\bf 794}, 87 (2007)
  [arXiv:hep-ph/0703025].
  \bibitem{prop}
  M. Yamada, 
  Phys. \ Rev. \ D {\bf 30}, 2144 (1984)
\bibitem{Sakai:2005yt}
  T.~Sakai and S.~Sugimoto,
  ``More on a holographic dual of QCD,''
  Prog.\ Theor.\ Phys.\ {\bf 114}, 1083 (2005)
  [arXiv:hep-th/0507073].
\bibitem{Brower:2000rp}
  R.~C.~Brower, S.~D.~Mathur and C.~I.~Tan,
  ``Glueball Spectrum for QCD from AdS Supergravity Duality,''
  Nucl.\ Phys.\  B {\bf 587}, 249 (2000)
  [arXiv:hep-th/0003115].
  %
\bibitem{Constable:1999gb}
  N.~R.~Constable and R.~C.~Myers,
  ``Spin-two glueballs, positive energy theorems and the AdS/CFT
  correspondence,''
  JHEP {\bf 9910}, 037 (1999)
  [arXiv:hep-th/9908175].
\bibitem{Adkins:1983ya}
  G.~S.~Adkins, C.~R.~Nappi and E.~Witten,
  ``Static Properties Of Nucleons In The Skyrme Model,''
  Nucl.\ Phys.\  B {\bf 228}, 552 (1983).
%
\bibitem{Skyrme:1961vq}
  T.~H.~R.~Skyrme,
  ``A Nonlinear field theory,''
  Proc.\ Roy.\ Soc.\ Lond.\  A {\bf 260}, 127 (1961).
\bibitem{Adkins:1983nw}
  G.~S.~Adkins and C.~R.~Nappi,
  ``Stabilization Of Chiral Solitons Via Vector Mesons,''
  Phys.\ Lett.\  B {\bf 137}, 251 (1984).
\bibitem{Morningstar:1999rf}
  C.~J.~Morningstar and M.~J.~Peardon,
  ``The glueball spectrum from an anisotropic lattice study,''
  Phys.\ Rev.\  D {\bf 60}, 034509 (1999)
  [arXiv:hep-lat/9901004].


\bibitem{Amos:1985wx}
  N.~A.~Amos {\it et al.},
  ``Measurement of Small Angle $\bar{p}p$ and Proton Proton Elastic Scattering
  at the CERN Intersecting Storage Rings,''
  Nucl.\ Phys.\  B {\bf 262}, 689 (1985).
\bibitem{Bernard:1987vq}
  D.~1.~Bernard {\it et al.}  [UA4 Collaboration],
  ``The Real Part of the Proton - anti-Proton Elastic Scattering Amplitude at
  the Center-Of-Mass Energy of 546-GeV,''
  Phys.\ Lett.\  B {\bf 198}, 583 (1987).
\bibitem{Donnachie:1983hf}
  A.~Donnachie and P.~V.~Landshoff,
  ``P P And Anti-P P Elastic Scattering,''
  Nucl.\ Phys.\  B {\bf 231}, 189 (1984).
\bibitem{Landshoff:1974db}
  P.~V.~Landshoff and J.~C.~Polkinghorne,
  ``Strong interactions at large transverse momentum. ii,''
  Phys.\ Rev.\  D {\bf 10}, 891 (1974).
\bibitem{Jaroszkiewicz:1974ep}
  G.~A.~Jaroszkiewicz and P.~V.~Landshoff,
  ``Model for diffraction excitation,''
  Phys.\ Rev.\  D {\bf 10}, 170 (1974).
\bibitem{DLbook}
A.~Donnachie, H.~G.~Dosch, P.~V.~Landshoff and O.~Nachtmann, {\it Pomeron Physics and QCD}, Cambridge 
University Press, Cambridge: 2002.
\bibitem{Ryskin:2009tj}
  M.~G.~Ryskin, A.~D.~Martin and V.~A.~Khoze,
  ``Soft processes at the LHC, I: Multi-component model,''
  Eur.\ Phys.\ J.\  C {\bf 60}, 249 (2009)
  [arXiv:0812.2407 [hep-ph]].
  \bibitem{Ryskin:2009tk}
  M.~G.~Ryskin, A.~D.~Martin and V.~A.~Khoze,
  ``Soft processes at the LHC, II: Soft-hard factorization breaking and gap
  survival,''
  Eur.\ Phys.\ J.\  C {\bf 60}, 265 (2009)
  [arXiv:0812.2413 [hep-ph]].
\bibitem{Bourrely:1984gi}
  C.~Bourrely, J.~Soffer and T.~T.~Wu,
  ``Impact Picture Expectations for Very High-Energy Elastic $p p$ and 
 $p \bar{p}$ Scattering,''
  Nucl.\ Phys.\  B {\bf 247}, 15 (1984).
  \bibitem{Block:1998hu}
  M.~M.~Block, E.~M.~Gregores, F.~Halzen and G.~Pancheri,
  ``Photon - proton and photon-photon scattering from nucleon-nucleon forward
  amplitudes,''
  Phys.\ Rev.\  D {\bf 60}, 054024 (1999)
  [arXiv:hep-ph/9809403].
\bibitem{Cudellpdg}
  J.~R.~Cudell {\it et al.},
  ``Hadronic scattering amplitudes: Medium-energy constraints on asymptotic
  behaviour,''
  Phys.\ Rev.\  D {\bf 65}, 074024 (2002)
  [arXiv:hep-ph/0107219].
\bibitem{Igipdg}
  K.~Igi and M.~Ishida,
  ``Investigations of the pi N total cross sections at high energies using  new
  FESR: log(nu) or (log(nu))**2,''
  Phys.\ Rev.\  D {\bf 66}, 034023 (2002)
  [arXiv:hep-ph/0202163].
\bibitem{Blockpdg}
  M.~M.~Block and F.~Halzen,
  ``Evidence for the saturation of the Froissart bound,''
  Phys.\ Rev.\  D {\bf 70}, 091901 (2004)
  [arXiv:hep-ph/0405174].
\bibitem{Hata:2007mb}
  H.~Hata, T.~Sakai, S.~Sugimoto and S.~Yamato,
  ``Baryons from instantons in holographic QCD,''
  arXiv:hep-th/0701280.
\bibitem{Hashimoto:2008zw}
  K.~Hashimoto, T.~Sakai and S.~Sugimoto,
  ``Holographic Baryons : Static Properties and Form Factors from Gauge/String
  Duality,''
  Prog.\ Theor.\ Phys.\  {\bf 120}, 1093 (2008)
  [arXiv:0806.3122 [hep-th]].
  %

\bibitem{Domokos:2010ma}
  S.~K.~Domokos, J.~A.~Harvey and N.~Mann,
  arXiv:1008.2963 [hep-th].

\bibitem{Hong:2007kx}
  D.~K.~Hong, M.~Rho, H.~U.~Yee and P.~Yi,
  ``Chiral dynamics of baryons from string theory,''
  Phys.\ Rev.\  D {\bf 76}, 061901 (2007)
  [arXiv:hep-th/0701276].
\bibitem{Hong:2007ay}
  D.~K.~Hong, M.~Rho, H.~U.~Yee and P.~Yi,
  ``Dynamics of Baryons from String Theory and Vector Dominance,''
  JHEP {\bf 0709}, 063 (2007)
  [arXiv:0705.2632 [hep-th]].
\bibitem{Hong:2007dq}
  D.~K.~Hong, M.~Rho, H.~U.~Yee and P.~Yi,
  ``Nucleon Form Factors and Hidden Symmetry in Holographic QCD,''
  Phys.\ Rev.\  D {\bf 77}, 014030 (2008)
  [arXiv:0710.4615 [hep-ph]].



\end{references}
\end{document}